\documentclass{jfm}
\usepackage{graphicx}
\usepackage{epstopdf,epsfig}
\usepackage[]{subcaption} 
\usepackage{multirow}
\usepackage{array}
\newcolumntype{x}[1]{>{\centering\arraybackslash\hspace{0pt}}p{#1}}
\usepackage{amsmath}
\usepackage{enumitem} 
\usepackage{placeins}
\usepackage{newtxtext}
\usepackage{newtxmath}
\usepackage{natbib}
\usepackage{hyperref}
\hypersetup{
    colorlinks = true,
    urlcolor   = blue,
    citecolor  = blue,
}
\usepackage{todonotes}

\newcommand{\RomanNumeralCaps}[1]
\linenumbers

\shorttitle{Wake dynamics of wind turbines in unsteady streamwise flow conditions}
\shortauthor{N. J. Wei, A. El Makdah, J. C. Hu, F. Kaiser, D. E. Rival, and J. O. Dabiri}

\title{Wake dynamics of wind turbines in unsteady streamwise flow conditions}

\author{Nathaniel J. Wei\aff{1,2,3}
  \corresp{\email{njwei@seas.upenn.edu}},
  Adnan El Makdah\aff{4},
  JiaCheng Hu\aff{4},
  Frieder Kaiser\aff{4},
  David E. Rival\aff{4,5},
 \and John O. Dabiri\aff{3,6}}

\affiliation{\aff{1}Mechanical Engineering and Applied Mechanics, University of Pennsylvania,
Philadelphia, PA 19104, USA
\aff{2}Andlinger Center for Energy and the Environment, Princeton University, Princeton, NJ 08544, USA
\aff{3}Graduate Aerospace Laboratories, California Institute of Technology, Pasadena, CA 91125, USA
\aff{4}Mechanical and Materials Engineering, Queen's University, Kingston, ON, Canada K7L 3N6
\aff{5}Institute of Fluid Mechanics, Technische Universit{\"a}t Braunschweig, 38108 Braunschweig, Germany 
\aff{6}Mechanical and Civil Engineering, California Institute of Technology, Pasadena, CA 91125, USA
}

\begin{document}

\maketitle

\begin{abstract}
The unsteady flow physics of wind-turbine wakes under dynamic forcing conditions are critical to the modeling and control of wind farms for optimal power density. Unsteady forcing in the streamwise direction may be generated by unsteady inflow conditions in the atmospheric boundary layer, dynamic induction control of the turbine, or streamwise surge motions of a floating offshore wind turbine due to floating-platform oscillations. This study seeks to identify the dominant flow mechanisms in unsteady wakes forced by a periodic upstream inflow condition. A theoretical framework for the problem is derived, which describes undulations in the wake radius and streamwise velocity that propagate as traveling waves downstream in the wake. These dynamics encourage the aggregation of tip vortices into large structures that are advected along in the wake. Flow measurements in the wake of a periodically surging turbine were obtained in an optically accessible towing-tank facility, with an average diameter-based Reynolds number of 300,000 and with surge-velocity amplitudes of up to 40\% of the mean inflow velocity. Qualitative agreement between trends in the measurements and model predictions is observed, supporting the validity of the theoretical analyses. The experiments also demonstrate large enhancements in the recovery of the wake relative to the steady-flow case, with wake-length reductions of up to 46.5\% and improvements in the available power at 10 diameters downstream of up to 15.7\%. These results provide fundamental insights into the dynamics of unsteady wakes and serve as additional evidence that unsteady fluid mechanics can be leveraged to increase the power density of wind farms.
\end{abstract}

\begin{keywords}
Aerodynamics, vortex dynamics, wakes
\end{keywords}

\section{Introduction}
\label{sec:intro}


As the deployment of renewable energy continues to gain momentum worldwide, it is increasingly necessary to consider ways to optimize the power-generation capacity of large aggregations of these energy-harvesting systems. For this task, increasing the power density, or power generated per unit occupied land or sea area, is a critical objective for maximizing the contribution of renewable-energy sources to global energy demands while minimizing space and infrastructure requirements. For wind turbines, a major limiting factor to the power density of a wind farm is the wake regions downstream of each turbine. Downstream turbines that operate in these regions of low-speed and highly turbulent flow suffer drastic losses in power generation, thereby decreasing the overall power density of the array on the order of ten to twenty percent \citep{barthelmie_modelling_2009}.

Several methods for reducing these wake losses have been explored in recent years, including layout optimization using wake models \citep[cf.][]{stevens_flow_2017}, wake steering by turbine yaw misalignment \citep[e.g.][]{howland_wind_2019}, and flow control \citep[cf.][]{meyers_wind_2022}. Many of these approaches operate under the assumption that the incoming flow and its interactions with the turbines in an array behave in steady or quasi-steady manners. By contrast, flow conditions in the atmospheric boundary layer are inherently unsteady, and fluctuations across a wide range of length and time scales affect the power generation and wake dynamics of real wind farms. Therefore, including the effects of unsteady dynamics in the design, optimization, and analysis of wind farms can potentially uncover new strategies for maximizing power density \citep{dabiri_theoretical_2020}.

Accordingly, significant efforts have recently been invested into leveraging unsteady fluid mechanics for improved turbine performance and reduced wake losses on downstream turbines. \cite{goit_optimal_2015}, \cite{munters_optimal_2017}, and \cite{munters_towards_2018} introduced the idea of dynamic induction control, in which oscillations in the thrust force of a turbine are generated to excite wake instabilities, increase the mixing of high-momentum fluid in the free-stream flow into the wake, and thereby achieve accelerated wake recovery relative to steady-flow turbine operation. Wind-tunnel experiments by \cite{frederik_periodic_2020} and \cite{van_der_hoek_experimental_2022} have demonstrated the effectiveness of this approach in enhancing the wake recovery downstream of a single turbine. A similar dynamic control strategy, involving individual pitch control of the turbine blades to create helical disturbances in the wake, has recently been developed by \cite{frederik_helix_2020} and tested in wind-tunnel experiments \citep{vander_hoek_maximizing_2024}. Other forms of dynamic wake control that introduce oscillations in the turbine rotation rate to excite tip-vortex pairing instabilities have been explored by \cite{brown_accelerated_2022}. \cite{meyers_wind_2022} provide a review of many recent studies of these kinds of wake-control approaches.

For floating offshore wind turbines (FOWTs), the possibility of periodic turbine motions as a function of unsteady wind gusts, wave forcing, and floating-platform hydrodynamics represents an additional opportunity for leveraging unsteady flows for increased power density in offshore wind farms. Building on the work of \cite{wen_power_2018}, \cite{johlas_floating_2021}, and others, \cite{wei_phase-averaged_2022} and \cite{wei_power-generation_2023} demonstrated using wind-tunnel experiments and analytical models that periodically surging turbines (i.e.\ turbines moving in linear oscillations along the direction of the incoming wind) can generate over six percent more power in certain loading conditions than equivalent stationary turbines. \cite{el_makdah_influence_2019} also measured large increases in power generation for turbines in axial ramp-up gusts, which could suggest that fixed-bottom turbines in unsteady streamwise flows can achieve time-averaged power-generation enhancements as well. The effects of unsteady turbine motions on the turbine wake and power generation of downstream turbines, however, are less well understood.

Several studies have investigated the wake dynamics of floating offshore wind turbines moving in streamwise rocking or linear-surge motions \citep[cf.][]{cioni_characteristics_2023}. \cite{fontanella_wind_2022} observed traveling-wave oscillations in the streamwise velocity downstream of a periodically surging wind turbine in a wind tunnel, but at a measurement distance of 2.3 turbine diameters into the wake, no changes in wake recovery were found. In wind-tunnel experiments, \cite{rockel_wake_2016} found that the rocking motions of a FOWT lead to a suppression in the entrainment of kinetic energy into the wake and therefore a decrease in the wake-recovery rate. Conversely, recent measurements by \cite{messmer_enhanced_2024} using a turbine mounted on an actuated Stewart platform showed increases in wake recovery in excess of 20\% relative to the fixed-turbine case, as a function of streamwise surge and transverse sway motions. Similarly, experiments by \cite{bossuyt_floating_2023} involving arrays of scaled FOWT models undergoing simultaneous wind and wave forcing demonstrated improvements in power density with increased wave-induced platform oscillations. Additionally, \cite{van_den_broek_optimal_2023} and \cite{van_den_berg_dynamic_2023} have investigated the coupled effects of dynamic induction control and turbine motions using free-vortex wake simulations, finding that improvements in wake recovery due to periodic turbine-thrust oscillations can in some cases be mitigated by the motions of the turbine, leading to a reduced overall effect on wake recovery. Based on the mixed results of these studies, it is apparent that a better understanding of the fluid mechanics underlying unsteady wake behaviors in FOWTs is needed.

The main purpose of the present work is to theoretically and experimentally investigate the unsteady wake dynamics generated by streamwise forcing from a turbine, in order to identify the dominant flow mechanisms and to determine their effects on wake dynamics and recovery. The current approach focuses on streamwise forcing because of the salience of these disturbances in FOWT platform dynamics \citep{johlas_large_2019,bossuyt_floating_2023}, the direct relevance to the dynamic induction control literature, and the potential for power-generation enhancements in streamwise unsteady flows \citep{dabiri_theoretical_2020,wei_power-generation_2023}. The theoretical analysis is designed to be agnostic to the source of wake unsteadiness, and is therefore equally applicable to stationary turbines with dynamic induction control and FOWTs undergoing streamwise surge motions. While the experiments presented in this work use a periodically surging turbine to generate unsteady wakes, the results and conclusions should in principle apply to dynamic induction control scenarios as well. Lastly, this study identifies fundamental flow mechanisms for unsteady wakes that may find broader applications outside of wind-energy contexts, including hydrokinetic turbines in tidal flows \citep[e.g.][]{scarlett_unsteady_2020}, streamwise-oscillating cylinders \citep[e.g.][]{currie_streamwise_1987}, bio-inspired propulsors in free- or intermittent-swimming conditions \citep[cf.][]{smits_undulatory_2019}, and vehicles propelled by oscillatory jets \citep[e.g.][]{ruiz_vortex-enhanced_2011}.

The work is structured as follows. In Section \ref{sec:theory}, a theoretical framework for wakes with oscillatory streamwise inflow conditions is described, and its implications for wake properties and vortex dynamics are discussed. The assumptions, strengths, and limitations of the analysis approach are also considered. In Section \ref{sec:methods}, an experiment to obtain flow measurements in the wake of a periodically surging turbine using an optical towing-tank facility is described. Results from the experiments are presented in Section \ref{sec:results} and are compared qualitatively with predictions from the theoretical model to demonstrate that the modeling approach captures the dominant physics of the unsteady-wake problem. Implications of the findings for wind-energy applications are surveyed at the end of Section \ref{sec:results}, and conclusions are given in Section \ref{sec:conclusions}.

\section{Theoretical considerations}\label{sec:theory}

In this section, we derive a system of coupled partial differential equations as a phenomenological model for the unsteady dynamics of a turbine wake with an oscillatory upstream boundary condition. This general formulation can be applied to dynamic induction control for stationary turbines and streamwise periodic surge motions in FOWTs. The approach relies on several simplifying abstractions from real turbine wakes, and is therefore not intended to be fully quantitative. However, the theoretical framework can still provide useful insights into the dominant flow physics of the unsteady-wake problem, and it can also be applied to a kinematic description of vortex dynamics in the near wake.

In this work, we denote time averages with overbars, phase averages with tildes, and amplitudes with circumflexes. Quantities referring to steady-flow or quasi-steady measurements are written with a zero subscript (e.g.\ $\lambda_0$). For velocities, lowercase letters represent velocity fields that vary in space and time, e.g.\ $u(x,r,t)$, where $r$ represents the radial coordinate. The uppercase letter $U$ denotes the radially averaged streamwise velocity, i.e.\ $u(r)$ averaged in space from $r=0$ to the wake radius $R$. The velocity of the turbine surge motions is defined as $\mathcal{U}(t)$ to distinguish it from these other velocities, and it is assumed to be periodic with the form

\begin{equation}
\frac{\mathcal{U}(t)}{U_\infty} = u^* \sin\left(2\pi \frac{t}{T}\right)\,,
    \label{eqn:surge_kinematics}
\end{equation}

\noindent where $T$ is the surge period and $u^*$ is the surge-velocity amplitude. The reduced frequency associated with these motions is 

\begin{equation}
    k = \frac{2\pi D}{TU_\infty}\,,
    \label{eqn:k}
\end{equation}

\noindent where $D$ is the turbine diameter. This is referred to as the Strouhal number in several recent studies on dynamic induction control for wind turbines \citep[e.g.][]{munters_towards_2018,frederik_periodic_2020,messmer_enhanced_2024}.

\subsection{Governing equations for streamwise-unsteady wake dynamics}
\label{sec:theory_eqns}

\begin{figure}
\centering
  \includegraphics[width=0.6\textwidth]{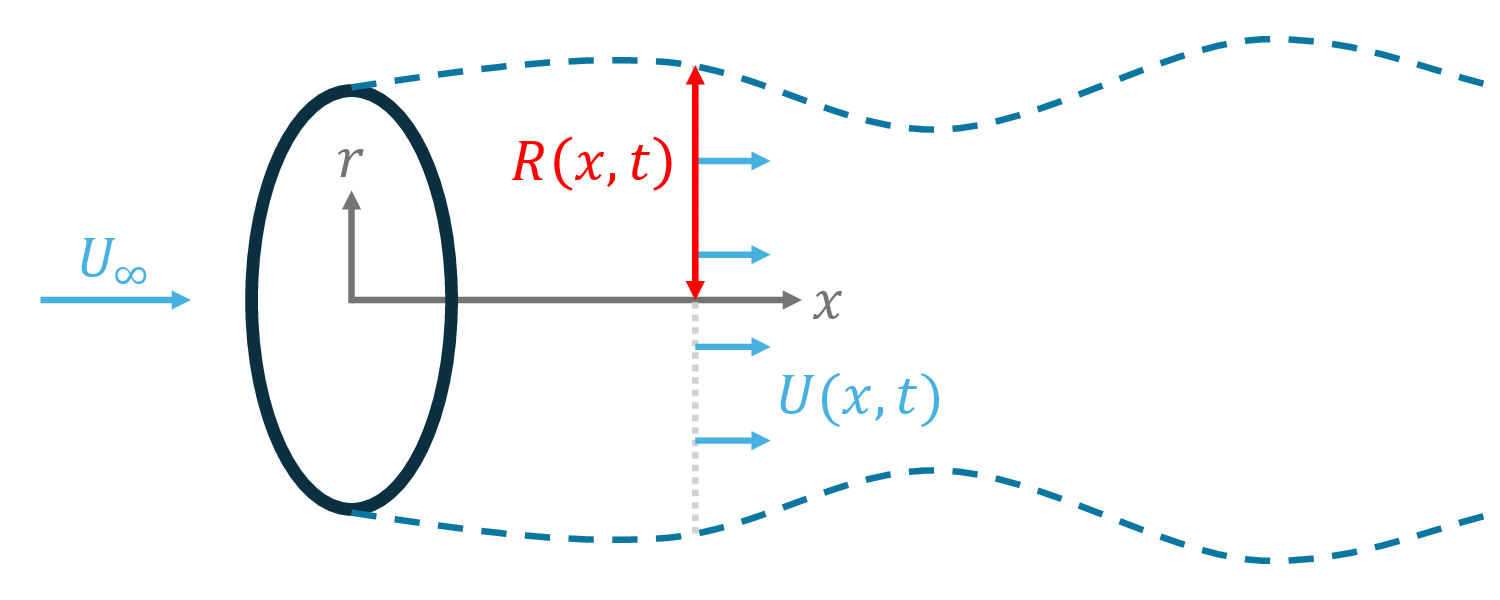}
  \caption{Sketch of the quasi-1D axisymmetric problem formulation for an unsteady turbine wake.}
\label{fig:CV}
\end{figure}

To derive the governing equations for a turbine wake with streamwise unsteadiness, we define a control volume with a variable wake radius $R(x,t)$, as shown in Figure \ref{fig:CV}. The flow inside the control volume is treated as incompressible and quasi one-dimensional, such that all flow occurs in the streamwise direction and the streamwise velocity $U(x,t)$ does not vary in the radial direction. The flow is additionally assumed to be inviscid in the near wake, where tip vortices dominate and viscous contributions to the kinetic-energy budget are comparatively small. The turbine itself is modeled as an actuator disc, which generates a thrust force on the flow and defines the inlet boundary condition to the control volume at $x=0$. Variations in the thrust force due to unsteady inflow conditions or streamwise surge motions will create an oscillatory inlet condition $U_i(x=0,t)$ that will dictate the dynamics in the near wake. We do not seek to model the coupling between the turbine dynamics and this near-wake inflow condition here, assuming for simplicity a periodic inflow condition of the form $U_i(x=0,t) = \overline{U} + \hat{U}\sin(ft)$.

Given these assumptions, the equations for conservation of mass and momentum within the wake control volume can be derived. (A detailed derivation is provided in Appendix \ref{app:derivations}.) The relation for conservation of mass is given as a first-order nonlinear partial differential equation in terms of $R$ and $U$:

\begin{equation}
    \frac{\partial R}{\partial t} + \frac{1}{2} R \frac{\partial U}{\partial x} + U \frac{\partial R}{\partial x} = 0\,.
    \label{eqn:continuity}
\end{equation}

\noindent If $r$ is held constant, Equation \ref{eqn:continuity} reduces to the more familiar form $\frac{\partial U}{\partial x} = 0$. The equation therefore captures the effects of changes in the flow velocity on the local size of the wake. For a quasi-1D incompressible flow, a decrease in the flow velocity across a given distance $\Delta x$ must result in a corresponding increase in the cross-sectional area of the control volume so that the mass flux remains constant. Similarly, an increase in the flow velocity will result in a decrease in the cross-sectional area. Hence, if $U(x,t)$ takes the form of a traveling wave, Equation \ref{eqn:continuity} dictates that the wake radius will also form traveling waves so that the mass flux at every streamwise location in the control volume is conserved.

The relation for conservation of momentum is the 1D Euler equation,

\begin{equation}
    \frac{\partial U}{\partial t} + U \frac{\partial U}{\partial x} = -\frac{1}{\rho} \frac{\partial p}{\partial x}\,.
    \label{eqn:momentum_Euler}
\end{equation}

\noindent Here, we have assumed that the pressure is constant in the radial direction, both inside and outside of the wake. The pressure term represents a heterogeneous forcing on a Burgers-type partial differential equation. In keeping with the assumptions of many turbine wake models \citep[e.g.][]{bastankhah_new_2014}, we choose to neglect this term, though in reality the pressure recovery in the near-wake region will be non-negligible.

For the remaining terms in the momentum equation, we decompose the flow velocity into phase-averaged and fluctuating components, in the style of a Reynolds decomposition:

\begin{equation}
U(x,t) = \tilde{U}(x,t) + U'(x,t)\,.
\end{equation}

\noindent Phase-averaging Equation \ref{eqn:momentum_Euler} using this decomposition yields

\begin{equation}
    \frac{\partial \tilde{U}}{\partial t} + \tilde{U} \frac{\partial \tilde{U}}{\partial x} = - \frac{1}{2} \frac{\partial}{\partial x} \widetilde{{U'}^2}\,.
    \label{eqn:momentum_RANS}
\end{equation}

To model the normal-stress term on the right-hand side of the equation, we combine two scaling relationships for wind-turbine wakes in steady inflow conditions. \citet{quarton_turbulence_1990} show that the turbulence intensity $u'/U_\infty$ decays monotonically toward zero with increasing streamwise distance. It is also well-known that the velocity in a turbine wake recovers monotonically toward the free-stream wind speed in a power-law relationship with streamwise distance \citep[cf.][]{porte-agel_wind-turbine_2020}. Thus, one can typically assume that the magnitude of the streamwise velocity gradient $\frac{\partial u}{\partial x}$ will decay monotonically toward zero as well. Combining these two observations, we argue that

\begin{equation}
    \widetilde{{U'}^2} \sim \left|\frac{d\tilde{U}}{dx}\right|\,,
\end{equation}

\noindent or in the form of a turbulent-viscosity hypothesis,

\begin{equation}
    -\widetilde{U'U'} = \nu_u \left|\frac{\partial \tilde{U}}{\partial x}\right|\,.
    \label{eqn:nu_T}
\end{equation}

\noindent The proportionality parameter $\nu_u$ may vary as a function of $x$; we will obtain a rudimentary model from experimental measurements in Section \ref{sec:results} of the form 

\begin{equation}
\nu_u = \kappa x 
\label{eqn:nu_T_model}
\end{equation}

\noindent (cf.\ Figure \ref{fig:nu_T}). It is important to note that $\nu_u$ is fundamentally different from a typical eddy viscosity based on the Reynolds shear stress, since it represents the interactions of streamwise quantities and has nothing to do with shear. Equation \ref{eqn:nu_T} thus suggests that stronger streamwise gradients are associated with stronger velocity fluctuations, and a linear model enforces a stronger coupling between the two quantities with increasing downstream distance. 


Applying this model to Equation \ref{eqn:momentum_RANS} yields

\begin{equation}
\frac{\partial \tilde{U}}{\partial t} + \tilde{U} \frac{\partial \tilde{U}}{\partial x} + \nu_u(x) \frac{\partial^2 \tilde{U}}{\partial x^2} = 0\,.
\label{eqn:momentum}
\end{equation}

\noindent This first-order nonlinear partial differential equation is a viscous Burgers equation, which describes the growth, propagation, and steepening of nonlinear traveling waves. The proportionality parameter $\nu_u(x)$ therefore represents a damping term that inhibits the steepening of the waves and the formation of unphysical shock discontinuities, which would occur in the inviscid form of the Burgers equation ($\nu_u=0$). This is physically consistent with the underlying $\widetilde{U'U'}$ quantity that the damping term represents, which is related to turbulent convection and transport in the turbulent kinetic energy budget for axisymmetric wakes \citep{uberoi_turbulent_1970}. This term thus models the transfer of energy from the phase-averaged base flow to turbulence via streamwise velocity gradients, which are created in this system by the time-varying inflow condition $U_i$, and the nonlinear wave steepening generated by the first two terms in Equation \ref{eqn:momentum} (i.e.\ the inviscid Burgers equation).

In summary, the dynamics of the unsteady wake can be written as a system of two first-order nonlinear partial differential equations,

\begin{subequations}
\label{eqn:model}
\begin{align}
    &\frac{\partial R}{\partial t} + \frac{1}{2} R \frac{\partial U}{\partial x} + U \frac{\partial R}{\partial x} = 0\;\rm{and} \label{eqn:modelR}\\
    &\frac{\partial U}{\partial t} + U \frac{\partial U}{\partial x} + \nu_u(x) \frac{\partial^2 U}{\partial x^2} = 0\,, \label{eqn:modelU}
\end{align}
\end{subequations}

\noindent where the tildes denoting phase-averaging have been removed for clarity and will not be used for the remainder of this work. The system is subject to the boundary conditions 

\begin{subequations}
\label{eqn:BCs}
\begin{align}
    \left. \frac{\partial R_i}{\partial t}\right|_{x=0,\,t} &= 0\;\rm{and} \label{eqn:BCR}\\
    \left. U_i\right|_{x=0,\,t} &=\overline{U}+\hat{U}\sin(ft)\,. \label{eqn:BCU}
\end{align}
\end{subequations}

\noindent The equations exhibit one-way coupling, as the momentum equation only depends on $U$ and serves as a forcing on $R$ through the continuity equation. The time-varying boundary condition $U_i$ and damping term in Equation \ref{eqn:modelU} preclude the straightforward derivation of explicit analytical solutions, but it is well-known that the viscous Burgers equation generates damped nonlinear traveling waves and can be solved numerically. For Equation \ref{eqn:modelR}, the method of characteristics can be applied by defining a characteristic variable $\xi = x - Ut$, such that along lines of constant $\xi$, the partial differential equation reduces to a pair of ordinary differential equations,

\begin{subequations}
\label{eqn:characteristics}
\begin{align}
    \left. \frac{dx}{dt}\right|_{\xi} &= U \;\rm{and} \label{eqn:char_x}\\
    \left. \frac{dR}{dt}\right|_{\xi} &= -\frac{1}{2}\frac{\partial U}{\partial x} R\,. \label{eqn:char_r}
\end{align}
\end{subequations}

\noindent The first equation describes the advection of solutions for $R$ according to the traveling-wave velocity $U$, while the second governs the growth or decay of the wake-radius amplitude independent of advection. Given the wave steepening that is inherent to solutions of Burgers-type equations for $U$, we expect that the magnitude of $\frac{\partial U}{\partial x}$ will be greater on the downstream side of the wave (where $\frac{\partial U}{\partial x}<0$) than on the upstream side of the wave. Thus, from Equation \ref{eqn:char_r}, we anticipate that the period-averaged amplitude of $R$ will grow under the forcing provided by Equation \ref{eqn:modelU} until streamwise velocity gradients are damped out, at which point the amplitude of $R$ will saturate and waves in the wake radius will simply advect downstream.

\begin{figure}
\begin{subfigure}[t]{0.48\textwidth}
\centering
  \includegraphics[width=\textwidth]{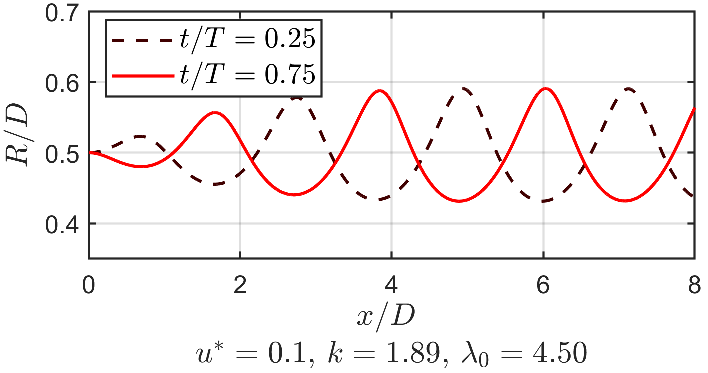}
  \caption{}
\label{fig:model_u1_r}
\end{subfigure}
\hfill
\begin{subfigure}[t]{0.48\textwidth}
\centering
  \includegraphics[width=\textwidth]{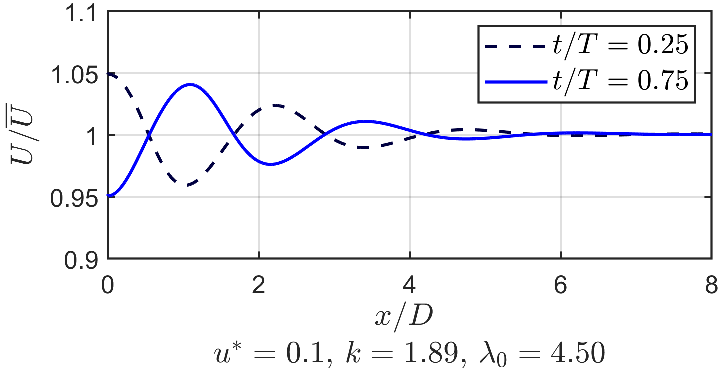}
  \caption{}
\label{fig:model_u1_u}
\end{subfigure}
\begin{subfigure}[t]{0.48\textwidth}
\centering
  \includegraphics[width=\textwidth]{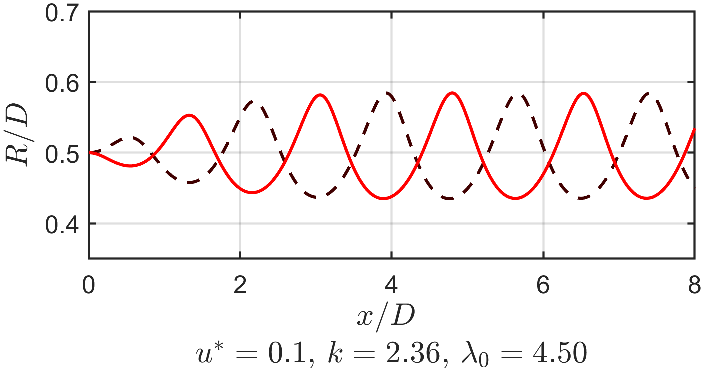}
  \caption{}
\label{fig:model_T08_r}
\end{subfigure}
\hfill
\begin{subfigure}[t]{0.48\textwidth}
\centering
  \includegraphics[width=\textwidth]{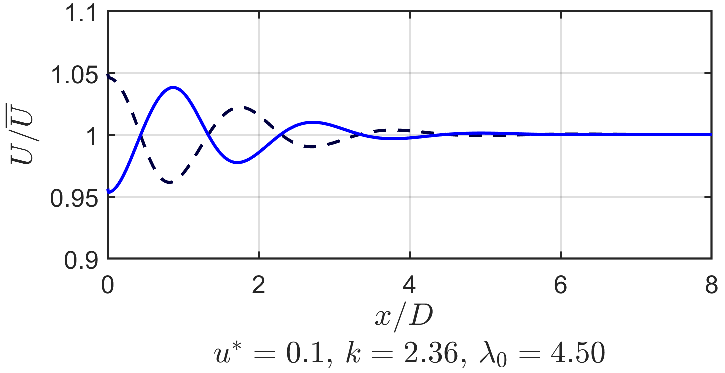}
  \caption{}
\label{fig:model_T08_u}
\end{subfigure}
\begin{subfigure}[t]{0.48\textwidth}
\centering
  \includegraphics[width=\textwidth]{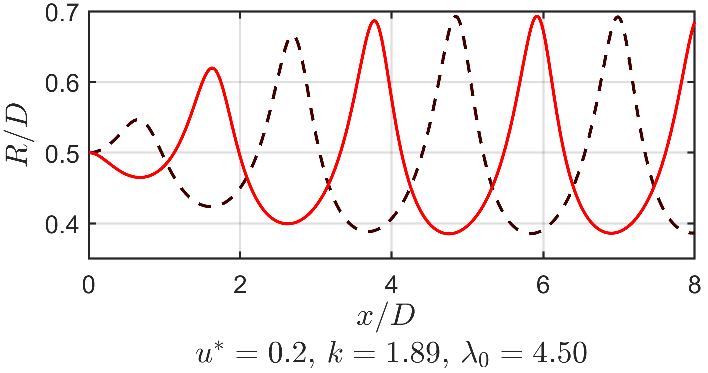}
  \caption{}
\label{fig:model_u2_r}
\end{subfigure}
\hfill
\begin{subfigure}[t]{0.48\textwidth}
\centering
  \includegraphics[width=\textwidth]{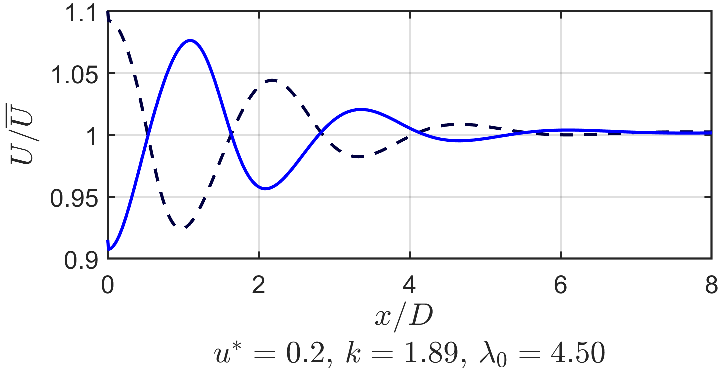}
  \caption{}
\label{fig:model_u2_u}
\end{subfigure}
\caption{Numerical solutions to the coupled-PDE modeling framework for the wake radius (left) and streamwise velocity (right), for three example cases.}
\label{fig:model_ru}
\end{figure}

To demonstrate the dynamics of the system, results from numerical integrations are shown in Figure \ref{fig:model_ru}. The upstream boundary conditions were set based on measured values from experiments (cf.\ Figure \ref{fig:ICs} in Section \ref{sec:results_dynamics}). A third-order upwind scheme was used to discretize the $\frac{\partial U}{\partial x}$ term, while a second-order central-difference scheme was used for the $\frac{\partial^2 U}{\partial x^2}$ term. As we have anticipated in our analysis, the model produces damped nonlinear traveling waves in the velocity, which in turn result in waves in the wake radius that grow and saturate in amplitude as they propagate downstream.

\subsection{Vortex dynamics}\label{sec:theory_vortex}

The kinematics of the control volume described by the system of equations derived in the previous section also allow predictions regarding the vortex dynamics in the wake to be made. Tip vortices shed by the turbine blades typically bound the wake of the turbine until they break down in the intermediate wake. This is a three-dimensional helical structure, but given the axisymmetric nature of the turbine wake, the successive appearances of the helical vortex line at a single azimuthal orientation are often treated as a discrete series of 2D point vortices \citep[e.g.][]{de_vaal_validation_2014,van_den_broek_optimal_2023}. In steady conditions, these vortex elements are generally arranged in relatively straight lines extending downstream from the blade tips. For an unsteady wake with a time-varying wake radius, the tip vortices will experience radial displacements along with the wake radius. We can therefore consider the dynamics of a series of discrete point vortices arranged in non-collinear patterns, as a representation of a helical tip vortex undergoing spatial and temporal changes in its radius. For more detailed stability analyses of helical tip vortices under dynamic conditions, we refer the reader to the work of \cite{kleine_stability_2022} on the wakes of moving FOWTs and \cite{rodriguez_stability_2021} on the wakes of turbines with flexible blades.

Consider an infinitely long line of point vortices with equal circulation $\Gamma$ and equal spacing $\Delta x$. The induced velocity from the $i$\textsuperscript{th} vortex on a given vortex $j$ is given by

\begin{equation}
    \begin{bmatrix}u\\v\end{bmatrix}_{i,j} = \frac{\Gamma}{2\pi\left((x_j-x_i)^2+(y_j-y_i)^2\right)}\begin{bmatrix} -(y_j - y_i) \\ x_j - x_i
    \end{bmatrix}\,,
    \label{eqn:vortexU}
\end{equation}

\noindent where we assume that radial perturbations ($\Delta r \equiv \Delta y$) are small such that $\Gamma$ is approximately constant. Due to the symmetries of the interactions, the induced velocity on a given vortex by its left neighbor will be canceled out by the induced velocity by its right neighbor, and the line will not deform over time.

Next, consider a similar system of point vortices with circulations $-\Gamma$, now arranged on a parabola of constant concavity given by $y=\alpha x^2$. For the vortex initially at $(x_j,y_j) = (0,0)$, the total induced velocity from its left and right neighbors at $j-1$ and $j+1$ is

\begin{equation}
    \begin{bmatrix}u\\v\end{bmatrix}_{j\pm1,j} = -\frac{\Gamma}{\pi\left(1+(\alpha \Delta x)^2\right)}\begin{bmatrix} \alpha \\ 0
    \end{bmatrix}\,.
\end{equation}


\noindent A point-vortex distribution with positive concavity will thus induce a negative tangential velocity in a given vortex, and the magnitude of this induced motion scales with the concavity $\alpha$. Conversely, a distribution with negative concavity will induce a positive tangential velocity in a vortex on the curve.

\begin{figure}
\centering
  \includegraphics[width=0.5\textwidth]{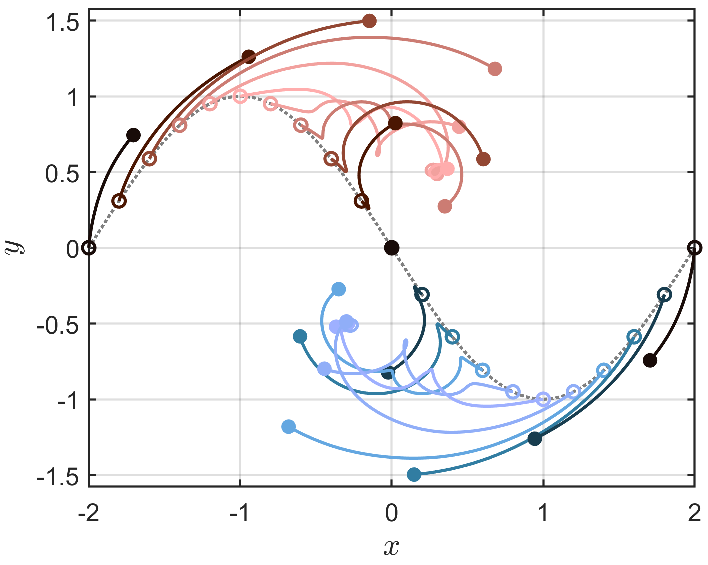}
  \caption{2D point-vortex simulation. Vortices with equal and negative (clockwise) circulation are initialized at the open circles on the grey dotted line. Their trajectories are shown in solid lines, and their final positions in the simulation are given by closed circles. Vortex locations and trajectories are colored by the concavity of the curve at the vortex's initial position.}
\label{fig:vortex_sim}
\end{figure}

This principle can be demonstrated for a system of point vortices initialized along a sine wave with no background flow and numerically integrated forward in time using Equation \ref{eqn:vortexU}, shown in Figure \ref{fig:vortex_sim}. Vortices initially located on the wave crest move in the positive direction, while vortices initially located on the wave trough move in the negative direction. In both cases, vortices traveling in the same direction are pushed into closer proximity with each other as they move, rolling up into a vortex aggregate.

Applying this analysis to the sample model solutions for the wake radius shown in Figure \ref{fig:model_ru}, we can predict the evolution of tip vortices in the turbine wake as they are advected downstream. Assuming the initial distribution of the tip vortices follows the model solutions for $r(x,t)$, we would expect to see a similar aggregation behavior as that observed in Figure \ref{fig:vortex_sim}. Furthermore, since the nonlinear steepening of the wake-radius waves leads to higher concavities at the wave crests relative to the wave troughs, the tip vortices should tend to aggregate most strongly downstream of the wave crests and ``surf'' on these waves as they travel downstream.

A complementary mechanism for tip-vortex aggregation can be found by considering the effect of streamwise-velocity gradients. Consider again a line of point vortices with initial spacing $\Delta x$, advected by a flow with a positive streamwise gradient $\frac{\Delta U}{\Delta x}$. If we follow one point vortex at its advection velocity $U$, after some time $\Delta t$, the distance between it and its neighbors will have increased to $\Delta x + \Delta U \Delta t$. By Equation \ref{eqn:vortexU}, this increase in separation will result in weaker interactions between neighboring vortices. Conversely, in flows with a negative streamwise gradient, the separation between vortices will decrease as a function of time, thereby increasing the magnitude of vortex interactions by mutual induction. 

Returning to the simulated traces of $R$ and $U$ in Figure \ref{fig:model_ru}, we observe that regions where $\frac{\partial U}{\partial x} < 0 $ coincide with peaks in the wake radius where the concavity of $R$ is most negative. The negative streamwise velocity gradients should therefore supplement the effect of negative concavity described previously and augment the tendency of tip-vortex elements to aggregate just downstream of peaks in the wake-radius waveform. By contrast, positive streamwise velocity gradients align with troughs in $R$, which will discourage the roll-up of individual vortices in these regions. These analyses suggest that tip vortices will primarily aggregate in a single structure just downstream of peaks in the wake-radius waves. The strength of this behavior will depend on the amplitude and frequency of the unsteady wake forcing $U_i(x=0,t)$, which dictate both the magnitude of the streamwise velocity gradients in the wake and the sharpness of the peaks in the wake-radius waveform.

The proposed mechanisms for vortex aggregation in unsteady wakes are similar in principle to that of tip-vortex pairing, a phenomenon that is related to the breakdown of turbine wakes in steady inflow conditions \citep{okulov_stability_2007,felli_mechanisms_2011,sarmast_mutual_2014,quaranta_long-wave_2015,lignarolo_tip-vortex_2015}. However, while the mutual-induction instabilities considered in canonical turbine wakes typically involve two or three vortex elements, the present analysis involves larger systems of vortices. The number of tip vortices shed into the wake over a single unsteady forcing period can be estimated as

\begin{equation}
    N_v = 2 N_b \frac{\lambda}{k}\,,
\end{equation}

\noindent where $N_b$ is the number of turbine blades, $\lambda$ is the tip-speed ratio, and $k$ is the reduced frequency. For the experiments that will be detailed in this work (cf.\ Table \ref{tab:unsteady} in Section \ref{sec:methods}), $11 < N_v < 33$. The unsteady vortex-interaction mechanisms described above occur across length scales on the order of half the wavelength of the wake-radius perturbations, thus involving about $N_v/2$ vortices per period. Therefore, while vortex-pairing instabilities may be present in periodically forced unsteady wakes, it is expected that the additional mechanisms described in this section will play a non-negligible role in the vortex dynamics of the wake, due to the number and distribution of the vortices present in the interactions.




In summary, the modeling approach outlined in this section provides an interpretable theoretical description of the dynamics of the wake radius, streamwise velocity, and tip-vortex aggregations in a periodically unsteady turbine wake. The framework offers insights into the dominant physics of the problem and can be used for qualitative predictions of trends in the dynamics, including that

\noindent\hangindent=12pt 1) Nonlinear traveling waves in the streamwise velocity $U$ are generated by the oscillatory boundary condition, undergo steepening, and are damped in amplitude as they advect downstream;

\noindent\hangindent=12pt 2) Traveling wakes in the wake radius $R$ are generated by the streamwise-velocity waves, grow with increasingly prominent crests and broad troughs, and saturate as velocity gradients dissipate;

\noindent\hangindent=12pt 3) Tip-vortex elements are encouraged by the unsteady wake dynamics to aggregate ahead of crests in the wake radius and advect downstream with these crests; and

\noindent\hangindent=12pt 4) Increasing the strength of the forcing in the boundary condition $U_i(x=0,t)$ increases the amplitudes of $U$ and $R$ in the wake, as well as the tendency toward vortex aggregation.

The modeling approach relies on several assumptions that limit its quantitative accuracy and restrict its applicability to the prediction of perturbations in the near wake. However, in the following sections we will demonstrate that it still captures many of the key dynamical features of unsteady turbine wakes.

\section{Experimental methods}\label{sec:methods}

\subsection{Experimental apparatus}\label{sec:methods_apparatus}

\begin{figure}
\centering
  \includegraphics[width=\textwidth]{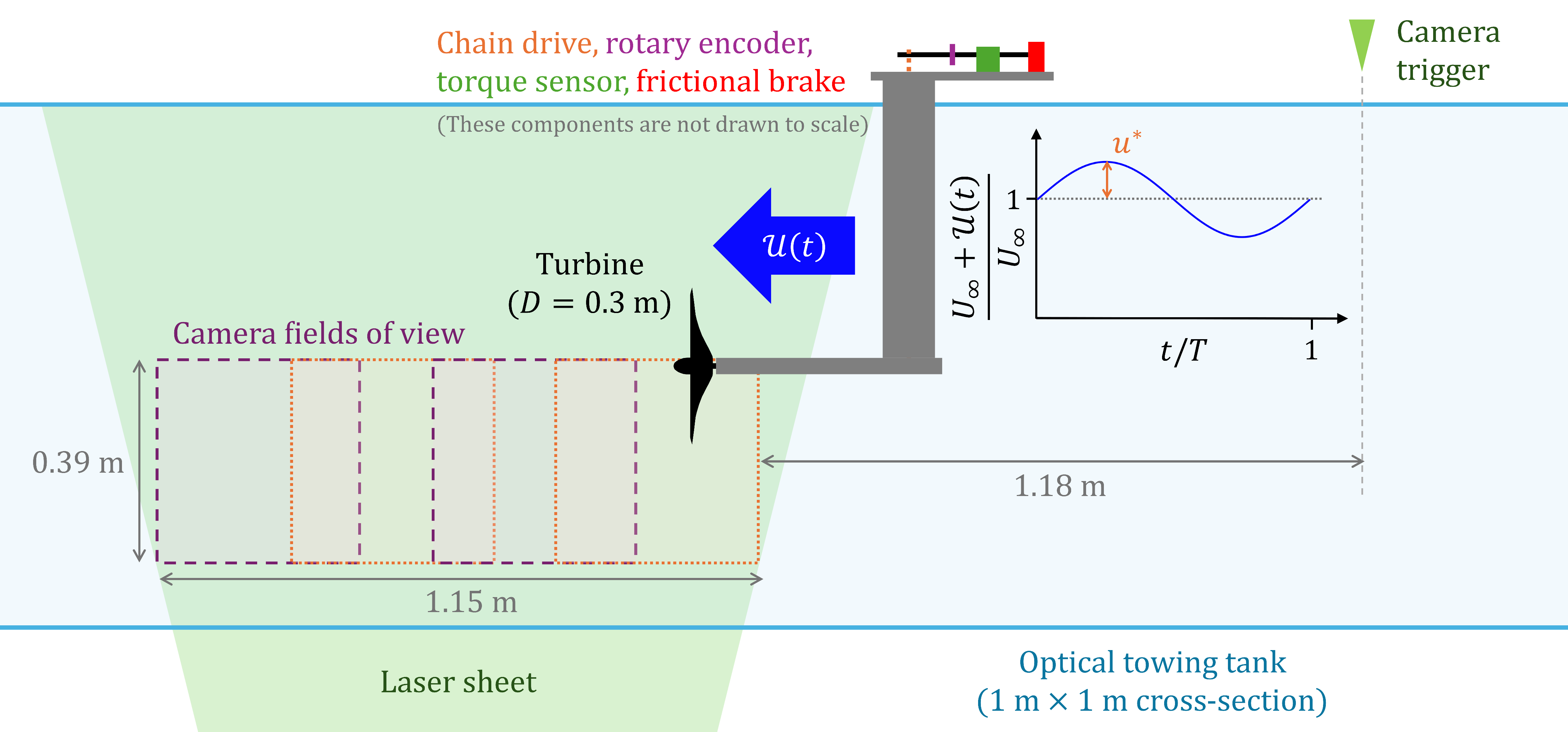}
  \caption{Schematic of the experimental apparatus in the optical towing tank, including the turbine, turbine sensors, traverse, and the fields of view of the four high-speed cameras. A sample velocity waveform $\mathcal{U}(t)$ for the traverse is shown in an inset.}
\label{fig:setup}
\end{figure}

Wake measurements downstream of a periodically surging rotor were conducted in an optically accessible towing-tank facility at Queen's University. The apparatus is described in detail by \cite{el_makdah_influence_2019}; a brief overview is given here, and a schematic is provided in Figure \ref{fig:setup}.

A 3D-printed turbine with a diameter of $D=0.3$ m was used in the experiments. The turbine blades were designed with SD7003 airfoil profiles of constant chord and a spanwise twist that targeted a constant angle of attack of $10^\circ$ along the blade at a tip-speed ratio of $\lambda=4$. The turbine had a maximum measured coefficient of power of $C_p=0.29$ at a tip-speed ratio of $\lambda = 3.89$. The turbine was mounted on a sting at the center of the towing-tank test section, and the turbine shaft was connected to a rotational encoder (Baumer ITD69H00), torque sensor (HBM T22), and frictional brake by means of a chain drive. The turbine system had a low rotational inertia relative to the hydrodynamic torques on the blades \citep{el_makdah_scaling_2021}. The estimated blockage of the turbine based on swept area was 7.1\%.

The test section of the towing tank was 15 m long and had a 1 m $\times$ 1 m cross-section. This was filled with water and enclosed by a ceiling that served to mitigate free-surface waves. The turbine was suspended from a traverse through a 50-mm wide opening in the ceiling. A rotary encoder on the traverse enabled its linear position to be recorded. The traverse was driven at a mean speed of $U_\infty=1$ $\rm{ms^{-1}}$ for steady-flow measurements. For unsteady surge cases, periodic oscillations in the traverse velocity of up to $\pm0.4$ $\rm{ms^{-1}}$ were superimposed on this mean speed. The average magnitude of the error between the desired velocity profile and the measured traverse velocity was 0.023 $\rm{ms^{-1}}$. Each traverse run was initiated with a constant-acceleration ramp-up profile that transitioned smoothly into the test motion waveform, and a similar ramp-down profile was used to bring the traverse to rest after the experiment. A light sensor mounted 3.84 m from the starting position of the traverse served as a trigger to start flow measurements. This position was chosen such that the traverse would reach steady-state operation before data recording was initiated. The towing tank was allowed to settle for at least four minutes between individual runs, as it was determined that longer settling durations did not further reduce the average particle displacements significantly. Each phase-averaged ensemble for a given case was compiled from 20 separate runs of the corresponding traverse-motion profile.

Four high-speed cameras (Photron SA4) with a resolution of $1024\times1024$ pixels were arranged in a line outside of the towing tank to capture a combined field of view of approximately 1.15 m $\times$ 0.39 m. The individual fields of view overlapped by approximately 30\%. The cameras were triggered simultaneously by the aforementioned light sensor and recorded at a frame rate of 500 Hz. A high-speed laser (Photonics DM40-527) was used to illuminate the measurement domain along the rotational axis of the turbine. The flow was seeded by neutrally buoyant polyamide spherical particles (LaVision) with a diameter of 60 $\rm{\mu m}$. The turbine and sting were spray-painted black to minimize reflections.

\subsection{Data analysis}\label{sec:methods_analysis}

2D particle-image velocimetry (PIV) was performed using the raw image data recorded by the cameras. An automated masking routine was written to mask the turbine, sting, and shadow cast by the turbine blades, based on the identified location of the nose of the turbine in the images. Only images recorded between the start trigger signal and the start of the traverse ramp-down motion were processed. Velocity vectors were computed using multi-pass cross-correlation with a final interrogation-window size of $32\times32$ pixels with 50\% overlap, using the open-source MATLAB package PIVlab \citep{thielicke_pivlab_2014}. A high-pass filter was applied to the images before correlation, and standard-deviation and median thresholds were applied to the vector fields after processing.

The individual vector fields from each camera were stitched together by interpolating onto a common spatial grid and averaging the velocities in the overlap regions. These composite lab-fixed velocity fields were then transferred into a reference frame moving at the mean speed of the turbine by identifying their locations in space and time within a single turbine surge period, interpolating onto a common spatial grid in the new reference frame, and stitching overlapping regions via linearly weighted blending. For the steady-flow reference cases, the representative period was set as $T=1$ s (for $U_\infty \leq 1$) or $T=0.5$ s (for $U_\infty > 1$). Vorticity fields were then calculated using Gaussian-filtered velocity fields to smooth out spurious results from numerical differentiation. Finally, the fields were phase-averaged across all 20 traverse runs, yielding time-resolved 2D velocity fields over a single period that covered at least 12 turbine diameters of the streamwise extent of the wake and over 1 turbine diameter in the radial direction. In some cases, over $18D$ of the wake was measured.

To serve as quantities for comparison with the flow model described in Section \ref{sec:theory}, the wake radius and streamwise velocity were also computed from the PIV data. The wake radius was defined at a given location $x$ as the radial location $r$ at which the streamwise velocity reached 95\% of the free-stream velocity, i.e.\ $R(x,t)$ such that $u(x,R,t) = 0.95U_\infty$. These profiles of $R(x,t)$ were smoothed using a moving-average filter with a width of 3 samples. The radially averaged streamwise velocity $U(x,t)$ was computed by averaging the local velocity $u$ across a streamwise slice of the wake at $x$, from the center of the wake to the wake radius $R(x,t)$.

\subsection{Experimental cases}\label{sec:methods_procedure}

The parameter space investigated in these experiments is summarized in Table \ref{tab:unsteady}. Two loading conditions were applied to the turbine. In the first case, the frictional brake was manually tuned between runs to obtain a steady-flow tip-speed ratio of $\lambda_0=4.50$. In the second case, the brake was released such that the only load on the turbine was due to shaft friction, yielding a steady-flow tip-speed ratio of $\lambda_0=5.07$. Both of these scenarios result in higher tip-speed ratios than the power-maximizing loading condition. This was intended to mitigate the effects of flow separation and stall on the turbine blades, in accordance with the observations of \cite{wei_phase-averaged_2022}. 

At each loading condition, both steady-reference and unsteady surge-motion cases were investigated. Steady-flow reference cases were carried out for each loading condition at a constant inflow velocity of $U_\infty = 1$ $\rm{ms^{-1}}$. Additional constant-velocity cases at $U_\infty = 0.8$ $\rm{ms^{-1}}$ and $U_\infty = 1.2$ $\rm{ms^{-1}}$ were collected to represent a quasi-steady range of wake profiles for a surge-velocity amplitude of $u^* = 0.2$, which will be compared against unsteady cases with the same surge-velocity amplitude to highlight differences between quasi-steady and unsteady wake dynamics.

Four unsteady cases for each loading condition spanned a range of surge-velocity amplitudes $u^*$ and reduced frequencies $k$: a baseline unsteady case with $u^*=0.1$ and $k=1.89$, a case with the same $u^*$ and higher $k$, a case with the same $k$ as the baseline case and higher $u^*$, and finally a case with as high of a value of $u^*$ as could be achieved with the apparatus, which required $k$ to be halved. For the case with $u^* = 0.4$, 20 additional runs were carried out in which the turbine motion waveform was offset by 1 m from its usual starting location, so that the combined set of 40 runs covered the entire wake over a single surge period. The surge-velocity amplitudes and reduced frequencies investigated in this study are relatively high and represent significant motions when applied to full-scale FOWTs, which in certain conditions experience motions with $u^*>0.25$ \citep{wayman_coupled_2006,larsen_method_2007,de_vaal_effect_2014}. The range of reduced frequencies is somewhat higher than the range expected for FOWTs as a function of wave and platform frequencies, which \cite{messmer_enhanced_2024} estimate to be $k\in[0,1.5]$. This was unavoidable due to limitations in the traverse velocity and length of the towing tank. Still, based on the findings of \cite{messmer_enhanced_2024}, we expect the dynamics observed at higher reduced frequencies to apply to somewhat lower reduced frequencies ($k\gtrsim0.5$) as well.

\begin{table}
  \begin{center}
  \begin{tabular}{x{2cm} x{2cm} x{1cm} x{1cm} x{1cm} x{2cm} x{2cm}}
        $\lambda_0$ & $C_{p,0}$ & $T$ [s] & $u^*$ & $k$ & $\overline{\lambda}$ & $\overline{C_p}$ \\\hline
        \multirow{4}{*}{$4.50\pm0.193$} & \multirow{4}{*}{$0.144\pm0.045$} & 1 & 0.1 & 1.89 & $4.50\pm0.229$ & $0.144\pm0.051$ \\
        & & 0.8 & 0.1 & 2.36 & $4.51\pm0.226$ & $0.139\pm0.047$ \\
        & & 1 & 0.2 & 1.89 & $4.52\pm0.225$ & $0.142\pm0.050$ \\
        & & 2 & 0.4 & 0.94 & $4.25\pm0.495$ & $0.123\pm0.043$ \\\hline
        \multirow{4}{*}{$5.07\pm0.173$} & \multirow{4}{*}{$0.033\pm0.025$} & 1 & 0.1 & 1.89 & $5.05\pm0.196$ & $0.034\pm0.024$ \\
        & & 0.8 & 0.1 & 2.36 & $5.05\pm0.207$ & $0.035\pm0.025$ \\
        & & 1 & 0.2 & 1.89 & $5.06\pm0.180$ & $0.034\pm0.022$ \\
        & & 2 & 0.4 & 0.94 & $5.04\pm0.185$ & $0.032\pm0.020$ \\
  \end{tabular}
  \caption{Operational parameters for the turbine used in these experiments. Two loading conditions were applied to the turbine, and the steady-flow tip-speed ratio $\lambda_0$ and coefficient of power $C_{p,0}$ are given as reference conditions in the leftmost two columns. For the unsteady cases, surge-motion parameters (including the period $T$ in seconds, surge-velocity amplitude $u^*$, and reduced frequency $k$) and the resulting time-averaged values of $\lambda$ and $C_p$ are given on the right side of the table. For all cases, $U_\infty = 1$ $\rm{ms^{-1}}$.}
  \label{tab:unsteady}
  \end{center}
\end{table}

\begin{figure}
\begin{subfigure}[t]{0.48\textwidth}
\centering
  \includegraphics[width=\textwidth]{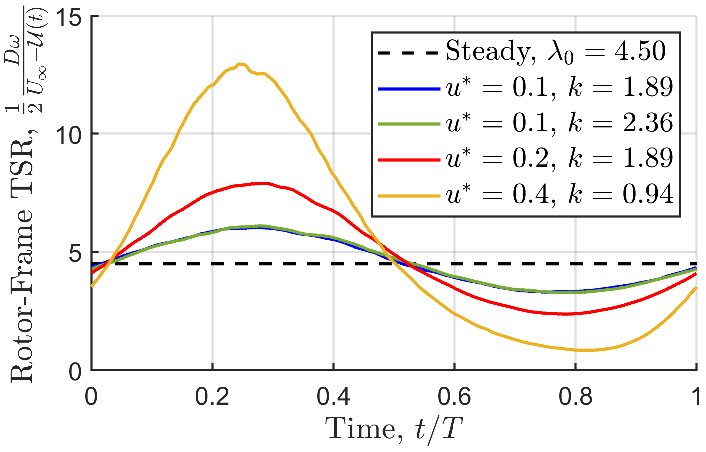}
  \caption{}
\label{fig:TSR_lo}
\end{subfigure}
\hfill
\begin{subfigure}[t]{0.48\textwidth}
\centering
  \includegraphics[width=\textwidth]{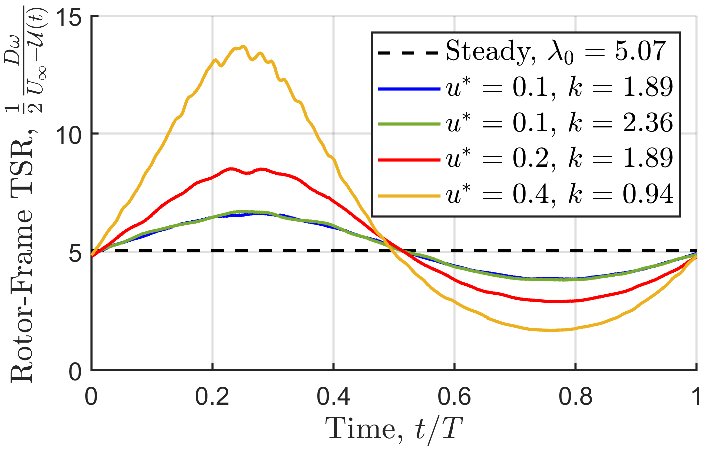}
  \caption{}
\label{fig:TSR_hi}
\end{subfigure}
\caption{Phase-averaged tip-speed ratio data, defined by the inflow velocity in the rotor frame, as a function of time for the two loading conditions covered in this study (a, b). The steady reference values are shown as black dashed lines.}
\label{fig:TSRs}
\end{figure}

The phase-averaged tip-speed ratios for all unsteady cases, referenced to the apparent inflow velocity in the rotor frame $U_\infty - \mathcal{U}(t)$, are shown in Figure \ref{fig:TSRs}. The fact that these signals are almost perfectly in phase with the surge-velocity waveform $\mathcal{U}(t)$ confirms that the inertia of the turbine is very low compared to the surge dynamics. We also note that, for most of the unsteady cases, the time-averaged tip-speed ratio and coefficient of power (shown in the rightmost columns of Table \ref{tab:unsteady}) remained relatively constant with changes in $u^*$ and $k$. For one case ($u^* = 0.4$ and $\lambda_0 = 4.50$), the time-averaged tip-speed ratio and power dropped due to the onset of stall in the turbine at low instantaneous inflow velocities, as evidenced by the drop in rotation rate observed around $t/T=0.8$ for this case in Figure \ref{fig:TSR_lo}.

\section{Results}\label{sec:results}

In this section, velocity and vorticity fields from the towing-tank experiments are shown to highlight the dominant features of the unsteady turbine wake. These dynamics are compared with the qualitative predictions of the modeling framework from Section \ref{sec:theory_eqns} to connect the previously discussed physical mechanisms with the observations. Finally, connections between the unsteady dynamics in the wake and enhancements in wake recovery observed in the unsteady cases are explored.

\subsection{Velocity and vorticity fields}\label{sec:results_fields}

\begin{figure}
\centering
  \includegraphics[width=\textwidth]{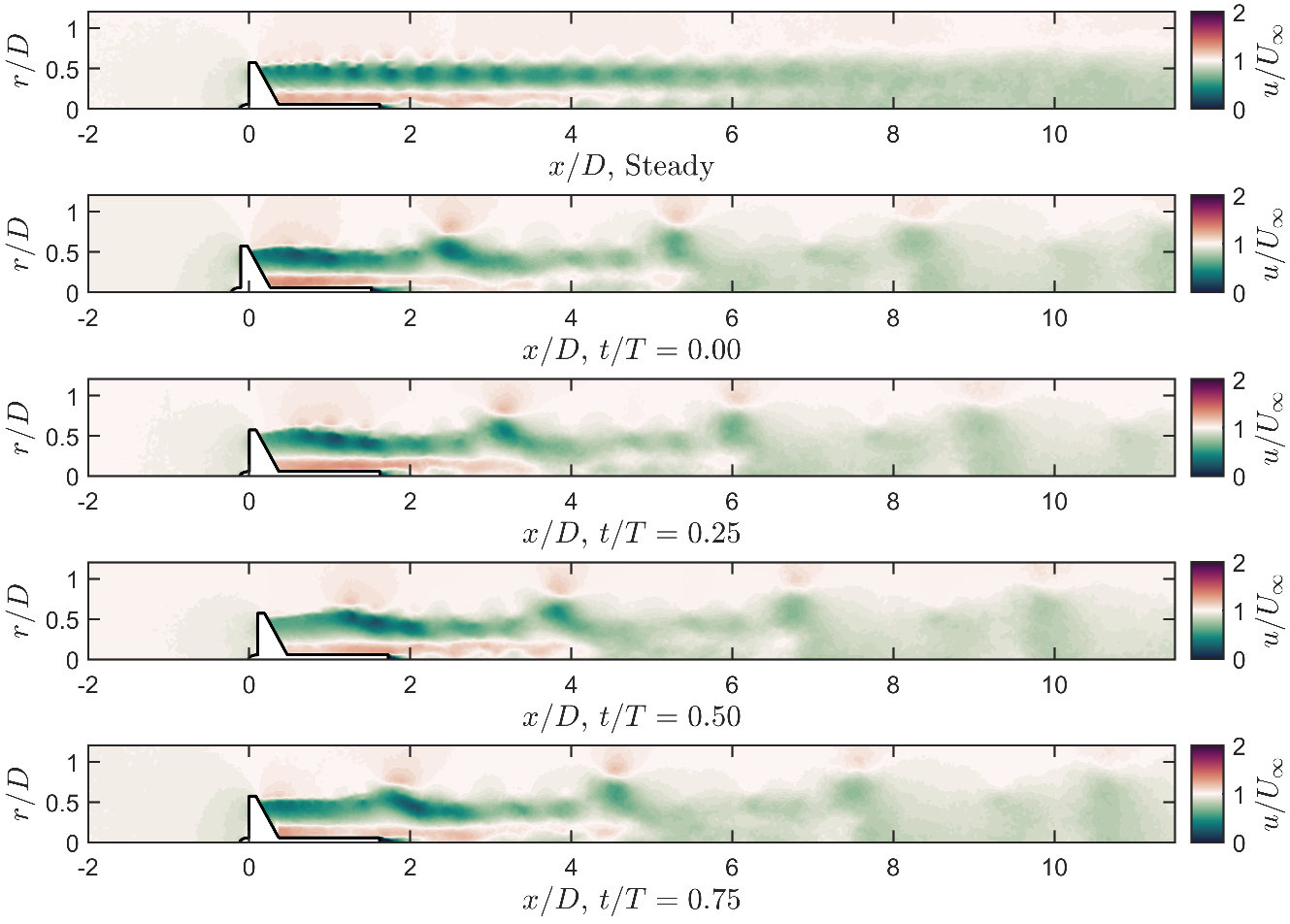}
  \caption{Streamwise-velocity fields for a steady-flow case (top) and four time-steps of an unsteady case with $u^* = 0.2$ and $k=1.89$. For both cases, $\lambda_0 = 5.07$.}
\label{fig:example_u}
\end{figure}

\begin{figure}
\centering
  \includegraphics[width=\textwidth]{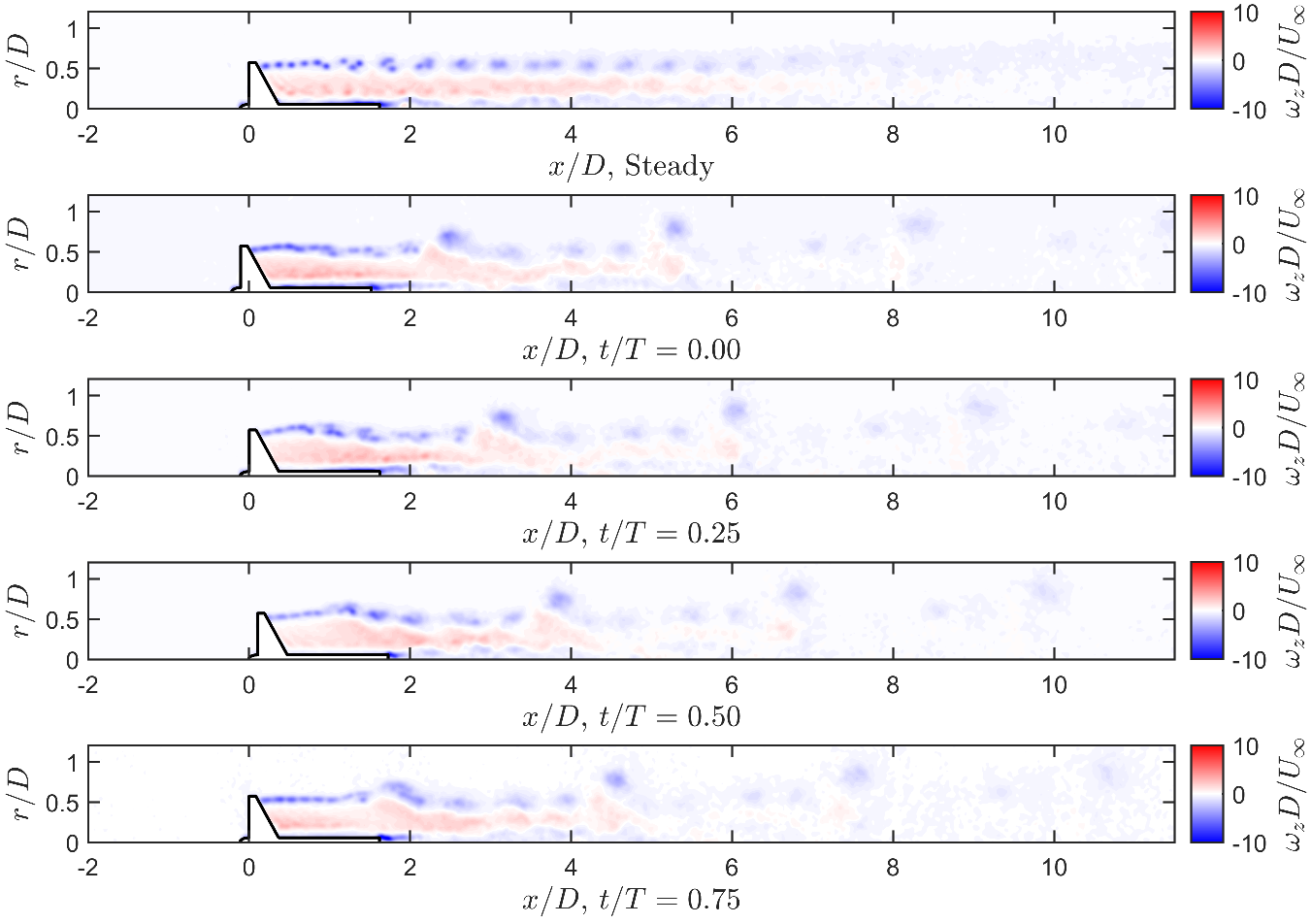}
  \caption{Out-of-plane vorticity fields for a steady-flow case (top) and four time-steps of an unsteady case with $u^* = 0.2$ and $k=1.89$. For both cases, $\lambda_0 = 5.07$.}
\label{fig:example_vort}
\end{figure}

First, to demonstrate key differences between the wakes from the steady-flow and unsteady cases, streamwise-velocity and out-of-plane vorticity fields are shown in Figures \ref{fig:example_u} and \ref{fig:example_vort} for sample cases at the higher reference tip-speed ratio, $\lambda_0 = 5.07$. The top plot in each figure shows a snapshot from the steady-flow wake, while the bottom four plots show four instantaneous snapshots from the unsteady wake ($u^*=0.2$ and $k=1.89$) at evenly spaced time steps. The steady wake shows typical wake features, such as gradual spreading in the velocity-deficit region, a monotonic recovery in the streamwise velocity with increasing downstream distance, and tip-vortex shedding in the near wake ($x/D \lesssim 3$) that breaks down in the intermediate wake. It is important to note that the azimuthal positions of the turbine blades were not synchronized across runs, so the fields shown in these and following figures are not phase-locked with respect to the turbine rotation. Despite this limitation, the signatures of tip vortices can still be identified in the vorticity fields. In the steady case shown in Figure \ref{fig:example_vort}, tip-vortex pairing instabilities are readily discernible, as the vortex elements collect into groups of three that then break down in the intermediate wake.

The unsteady wake, by contrast, exhibits strong departures from steady-flow wake behaviors. In Figure \ref{fig:example_u}, a pulsatile streamwise velocity is visible in the near wake around $x/D\approx 1$, as a result of the time-varying thrust and power of the turbine. This oscillatory inflow condition propagates downstream as a traveling wave and creates periodic peaks in the velocity-deficit region where the wake radius extends out past the steady-flow wake boundary. In the corresponding vorticity fields in Figure \ref{fig:example_vort}, the tip-vortex elements roll up into a larger aggregate structure at around $x/D\approx 2$, which is then advected downstream as a coherent vortex packet. While the signatures of tip-vortex pairing are still visible in these unsteady snapshots, the smaller aggregates that form on account of the mutual-inductance instability are subsumed into the larger aggregate structure. Additionally, the large vortex aggregate is shed with a periodicity matching that of the turbine surge waveform, suggesting that its dynamics are governed more strongly by the periodic wake forcing than by conventional vortex-pairing mechanisms. All of these observations align well with the theoretical analyses that were presented in Section \ref{sec:theory}, as well as the numerical simulations of \cite{kleine_stability_2022}.

\begin{figure}
\centering
  \includegraphics[width=\textwidth]{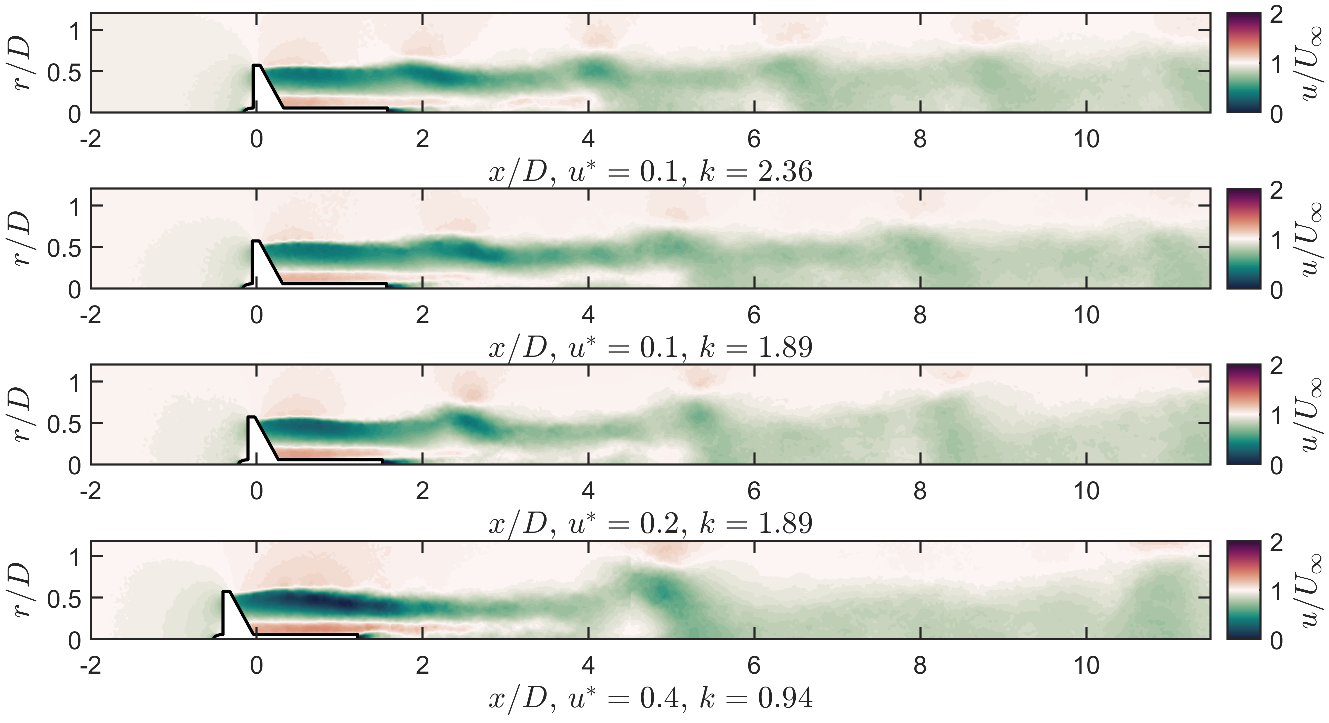}
  \caption{Streamwise-velocity fields for four unsteady cases at $t/T = 0$, with $T$ and $u^*$ increasing from top to bottom. All cases have $\lambda_0 = 4.50$.}
\label{fig:cases_u}
\end{figure}

\begin{figure}
\centering
  \includegraphics[width=\textwidth]{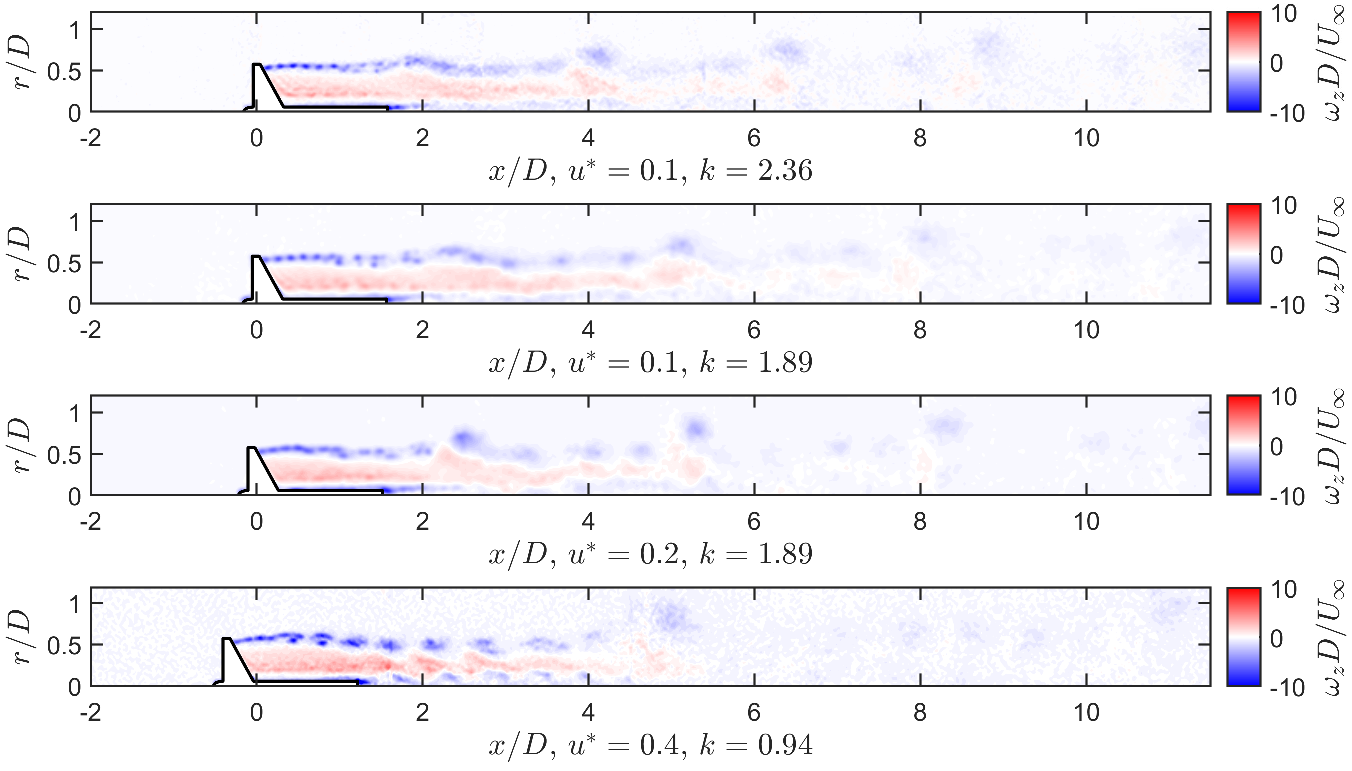}
  \caption{Vorticity fields for four unsteady cases at $t/T = 0$, with $T$ and $u^*$ increasing from top to bottom. All cases have $\lambda_0 = 5.07$.}
\label{fig:cases_vort}
\end{figure}

To demonstrate the effects of varying the unsteady surge-motion parameters $u^*$ and $k$, instantaneous snapshots at a fixed time instance $t/T = 0$ are shown for all four unsteady cases at the lower tip-speed ratio ($\lambda_0 = 4.50$) for the streamwise velocity (Figure \ref{fig:cases_u}) and out-of-plane vorticity (Figure \ref{fig:cases_vort}). The dynamics observed in the example case discussed above are also visible in these instances. The degree of unsteadiness in the wake appears to increase with increasing surge-velocity amplitude, while the wavelength of the traveling waves in the wake scales inversely with the reduced frequency $k$. In all cases, the traveling-wave dynamics and vortex aggregates appear to persist well into the far wake ($x/D\gtrsim 10$). The vortex aggregates that survive into the far wake are those that originally rolled up at the crests in the wake-radius wave, whereas no corresponding structures from the wake-radius wave troughs are visible past $x/D\approx8$. Again, these observations are well in accordance with the theoretical conjectures advanced in Section \ref{sec:theory}.

\subsection{Wake features and dynamics}\label{sec:results_dynamics}

\subsubsection{Trends in the wake radius and radially averaged streamwise velocity}

\begin{figure}
\centering
  \includegraphics[width=\textwidth]{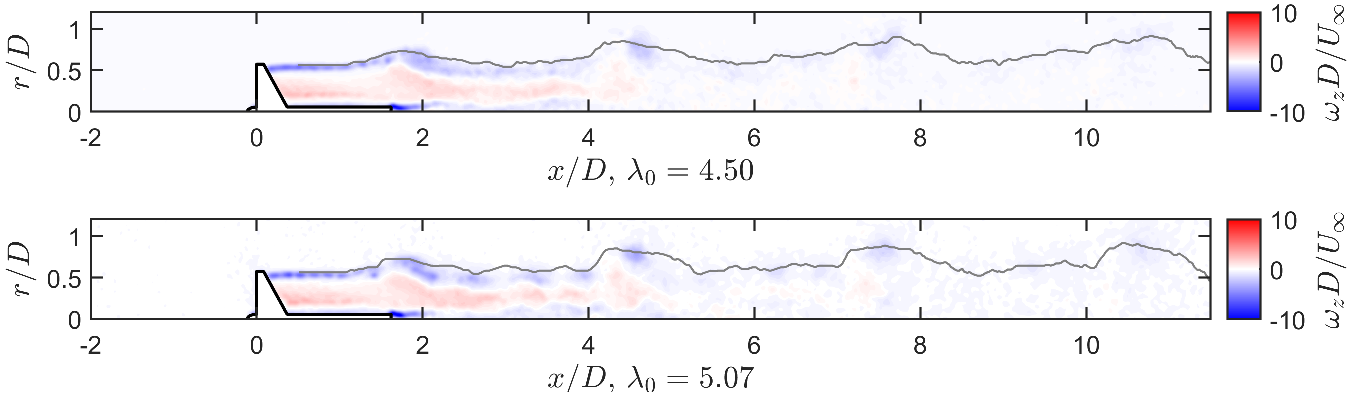}
  \caption{Calculated wake radii (gray line), superimposed on vorticity fields for two cases with $u^* = 0.2$ and $k=1.89$. The top and bottom plots show reference loading conditions of $\lambda_0 = 4.50$ and $\lambda_0 = 5.07$, respectively. Both cases are shown at $t/T = 0.75$.}
\label{fig:example_wakeRadius}
\end{figure}

To quantify the dynamics observed in the velocity and vorticity fields, we use the definitions of the wake radius and radially averaged streamwise velocity given in Section \ref{sec:methods_analysis} to produce experimental analogues to the model parameters $R$ and $U$ from Equations \ref{eqn:modelR} and \ref{eqn:modelU}. Examples of the calculated wake radius $R(x,t)$ are shown as grey lines on two vorticity snapshots in Figure \ref{fig:example_wakeRadius}. This visualization highlights the undulatory behavior of the wake in the unsteady cases.


\begin{figure}
\begin{subfigure}[t]{0.48\textwidth}
\centering
  \includegraphics[width=\textwidth]{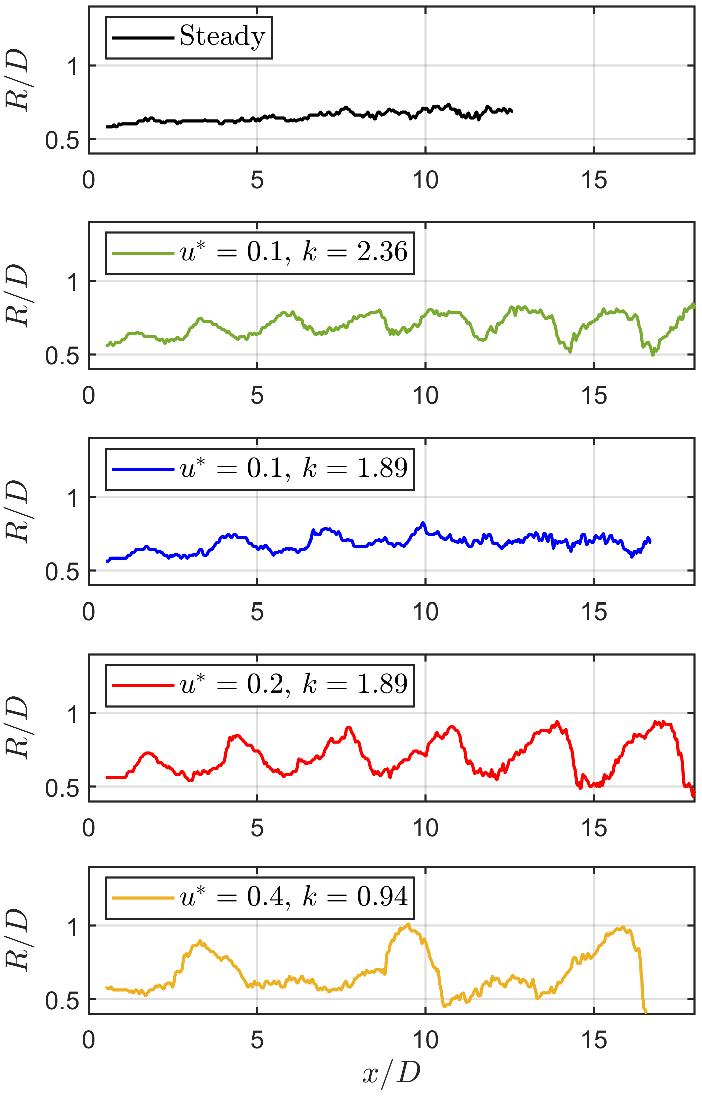}
  \caption{}
\label{fig:r_vs_x_data}
\end{subfigure}
\hfill
\begin{subfigure}[t]{0.48\textwidth}
\centering
  \includegraphics[width=\textwidth]{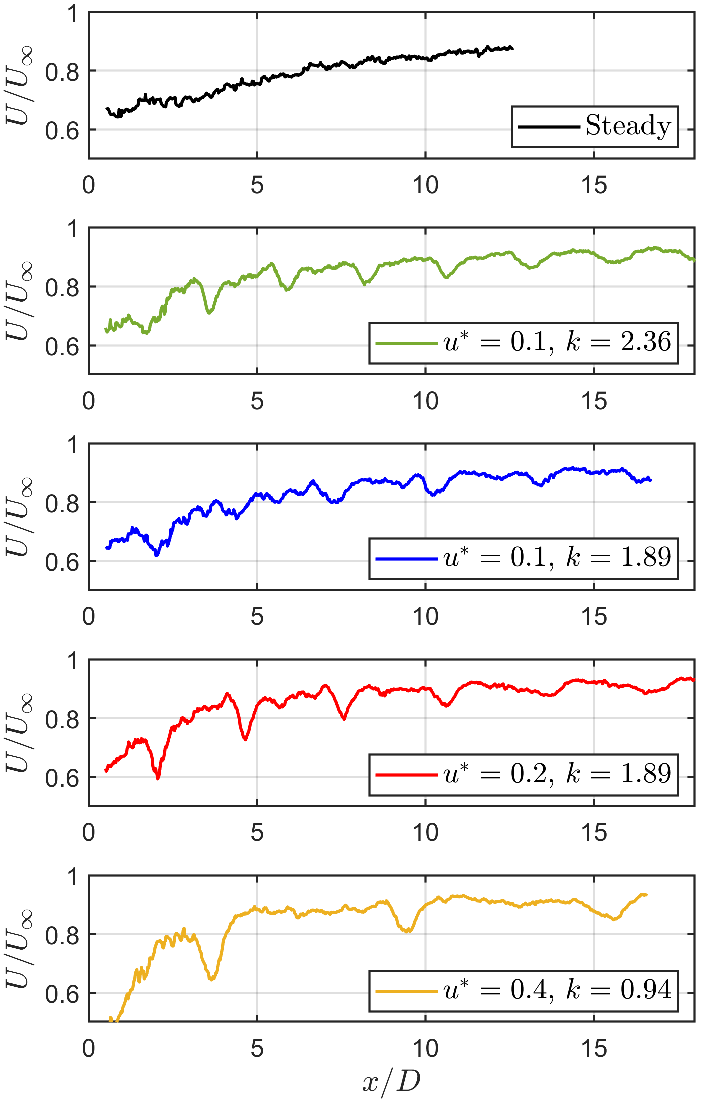}
  \caption{}
\label{fig:u_vs_x_data}
\end{subfigure}
\caption{Wake radius $R$ (a) and radially averaged streamwise velocity $U$ (b) as a function of streamwise distance at $t/T = 0.75$. All cases with $\lambda_0 = 4.50$ are shown.}
\label{fig:ru_vs_x_data}
\end{figure}

The calculated wake radius and radially averaged streamwise velocity are then shown for all of the steady and unsteady cases at $\lambda_0=4.50$ for a single time instance $t/T=0.75$ in Figure \ref{fig:ru_vs_x_data}. The wake-radius data in Figure \ref{fig:r_vs_x_data} show evidence of nonlinear traveling waves, with wave steepening on the downwind sides of the waves in the far wake for the higher-amplitude cases. Corresponding perturbations in the streamwise velocity are visible in Figure \ref{fig:u_vs_x_data}, and the magnitude of these perturbations decreases with increasing streamwise distance into the wake. These data confirm the salience of the dynamics captured by the modeling framework, but also demonstrate the limitations of the theoretical analysis. Both $R$ and $U$ demonstrate strong evidence of nonlinear traveling waves, as predicted by the model. However, the model assumes a constant time-averaged base-flow velocity $\overline{U}$ that does not change as a function of downstream distance. The data in Figure \ref{fig:u_vs_x_data}, by contrast, show almost immediate signs of wake recovery at relatively short downstream distances. This therefore limits the scope of the model to perturbations about time-averaged quantities. These considerations should be kept in mind as we now turn to compare the model with the data.

\subsubsection{Comparisons between model solutions and measured data}

While the modeling framework has limitations that may preclude its use for quantitatively accurate predictions, a comparison of its outputs with experimental data can still demonstrate that it is parameterizing the dominant physics of the unsteady-wake problem. To this end, we use the PIV data to extract representative inflow boundary conditions and a scaling for the turbulent fluctuations, and integrate the model with these parameters for direct comparison with the data.

\begin{figure}
\begin{subfigure}[t]{0.48\textwidth}
\centering
  \includegraphics[width=\textwidth]{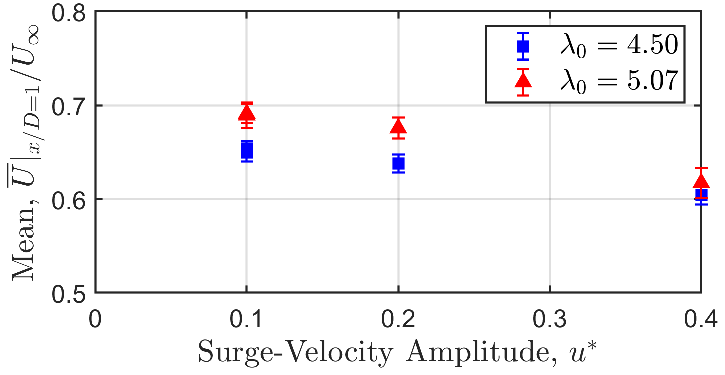}
  \caption{}
\label{fig:IC_uMean}
\end{subfigure}
\hfill
\begin{subfigure}[t]{0.48\textwidth}
\centering
  \includegraphics[width=\textwidth]{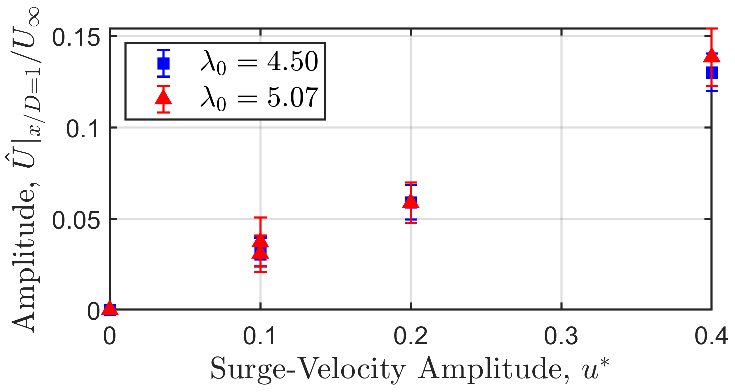}
  \caption{}
\label{fig:IC_uAmp}
\end{subfigure}
\caption{Time-averaged streamwise velocity (a) and amplitude of the streamwise velocity (b), both spatially averaged across the wake at $x/D = 1$. These data define the initial conditions for $U_i(x=0,t)$ in the analytical model, given in Equation \ref{eqn:BCU}.}
\label{fig:ICs}
\end{figure}

To define the upstream boundary condition $U_i(x=0,t)$ for the wake, the time average $\overline{U}$ and amplitude $\hat{U}$ of the radially averaged streamwise velocity were extracted from each unsteady dataset at $x/D = 1$ and are shown in Figure \ref{fig:ICs}. Notably, the amplitude of the streamwise velocity in the near wake appears to scale linearly with the surge-velocity amplitude in Figure \ref{fig:IC_uAmp}. These quantities were applied directly to the boundary condition in Equation \ref{eqn:BCU}.

\begin{figure}
\centering
  \includegraphics[width=\textwidth]{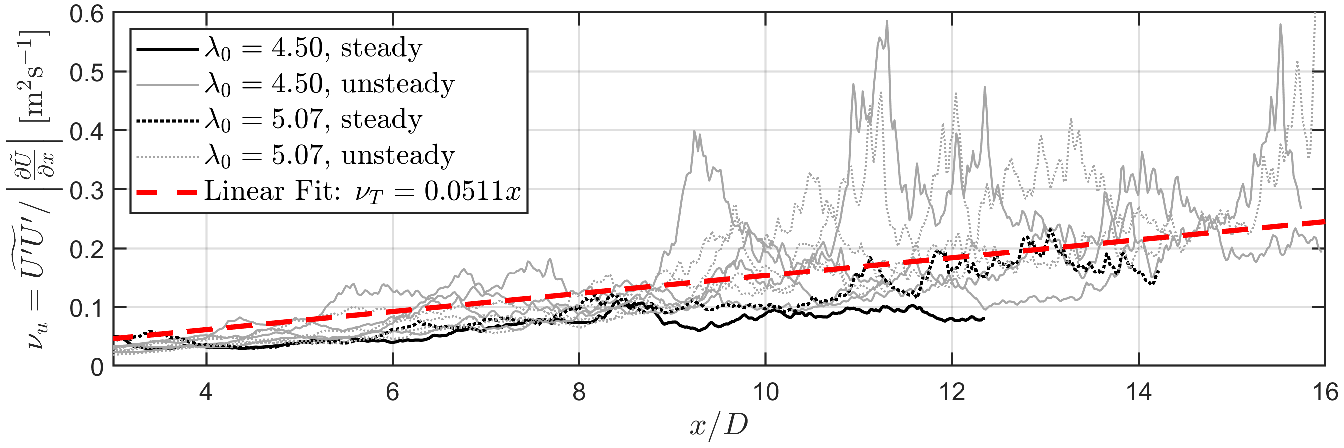}
  \caption{Empirical estimates of the proportionality parameter $\nu_u$, derived from Equation \ref{eqn:nu_T}. The fit to all of the data in the figure, shown as a red dashed line, was used as a rudimentary model for $\nu_u(x)$ in Equation \ref{eqn:modelU}. The fit best matches the data for $3 \lesssim x/D \lesssim 9$, whereas the data become noisier further downstream.}
\label{fig:nu_T}
\end{figure}

To estimate the coupling between streamwise-velocity fluctuations and streamwise-velocity gradients, as modeled by Equation \ref{eqn:nu_T}, the variances of the individual streamwise velocities $u'(x,r,t)$ across all 20 ensembles were computed. These values were then averaged across the wake at every streamwise location to estimate the streamwise-velocity autocovariance $\widetilde{U'U'}$ at each location. Dividing these quantities by the magnitude of the velocity gradient at each location gave estimates for the proportionality parameter $\nu_u(x)$. These empirical values are shown in Figure \ref{fig:nu_T}, and since they were observed to scale approximately linearly with streamwise distance in the intermediate wake ($3\lesssim x/D \lesssim 9$), a linear fit (cf.\ Equation \ref{eqn:nu_T_model}) over all of the steady and unsteady estimates was employed as the model for $\nu_u(x)$ in Equation \ref{eqn:modelU}. The applicability of the linear fit breaks down in the far wake ($x/D \gtrsim 9$), as the streamwise velocity gradients decrease in magnitude and the estimates for $\nu_u$ thus become more susceptible to noise. The overarching theoretical framework is not expected to apply in this region anyway since it neglects wake-recovery effects due to turbulent momentum entrainment, so the linear approximation was deemed acceptable for the purposes of this study. With this model for $\nu_u$ in place, Equations \ref{eqn:modelR} and \ref{eqn:modelU} could then be integrated numerically, as described in Section \ref{sec:theory_eqns}.

\begin{figure}
\begin{subfigure}[t]{0.48\textwidth}
\centering
  \includegraphics[width=\textwidth]{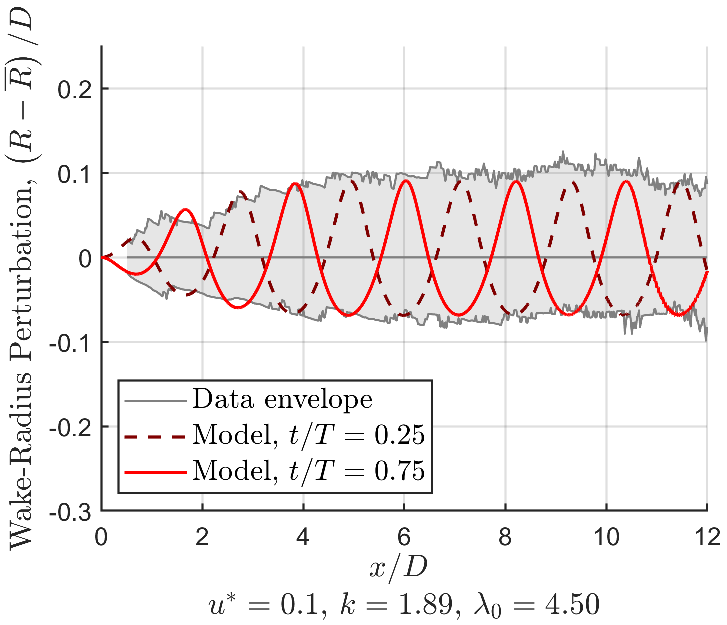}
  \caption{}
\label{fig:envelope_u1}
\end{subfigure}
\hfill
\begin{subfigure}[t]{0.48\textwidth}
\centering
  \includegraphics[width=\textwidth]{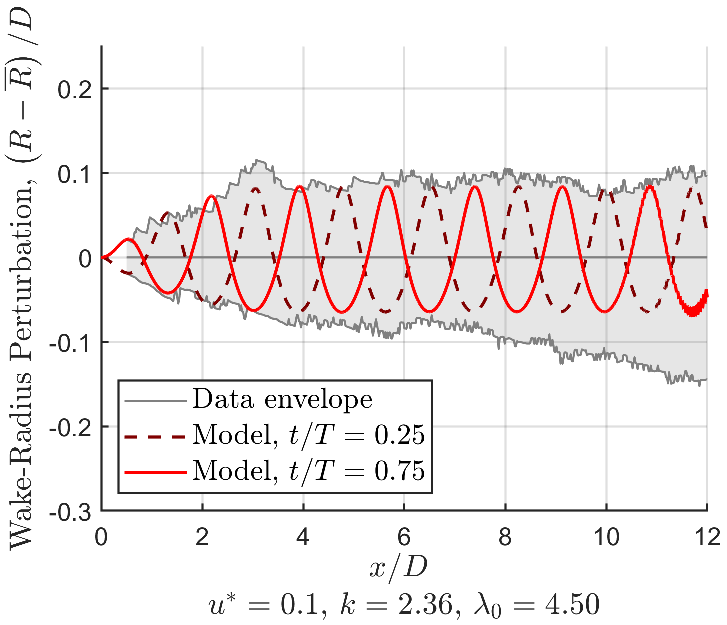}
  \caption{}
\label{fig:envelope_T08}
\end{subfigure}
\begin{subfigure}[t]{0.48\textwidth}
\centering
  \includegraphics[width=\textwidth]{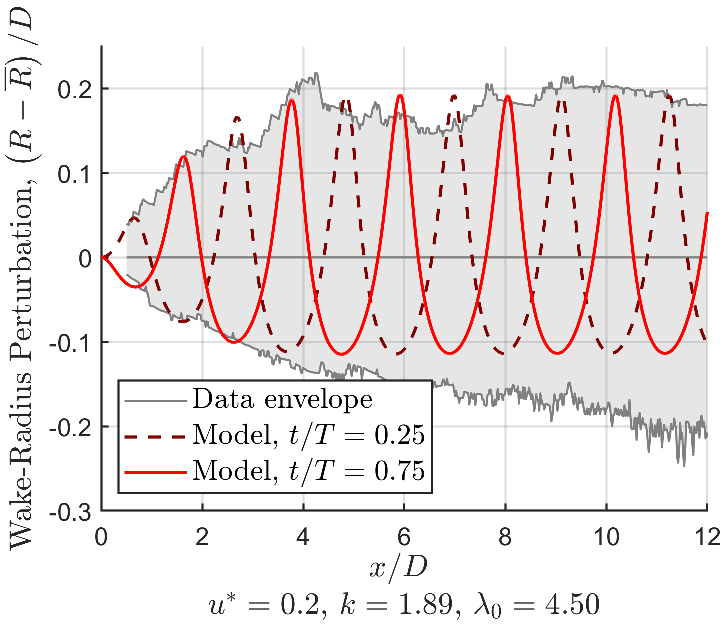}
  \caption{}
\label{fig:envelope_u2}
\end{subfigure}
\hfill
\begin{subfigure}[t]{0.48\textwidth}
\centering
  \includegraphics[width=\textwidth]{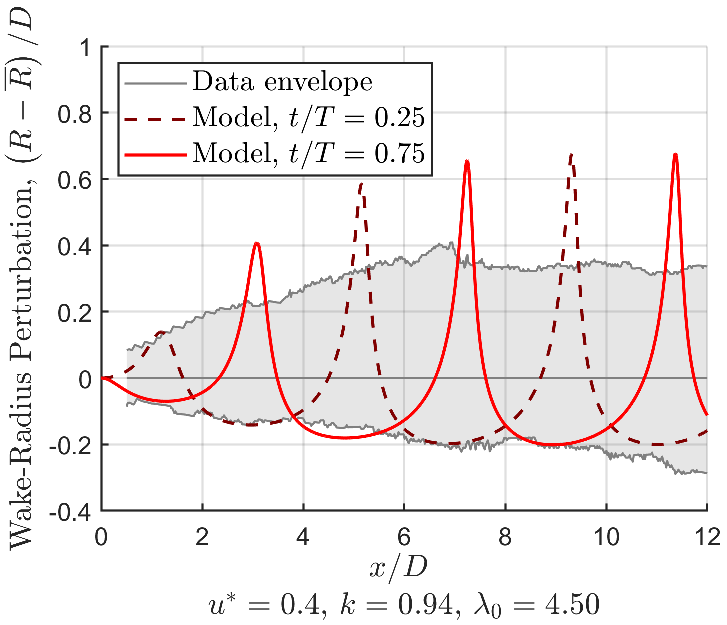}
  \caption{}
\label{fig:envelope_u4}
\end{subfigure}
\caption{Excursions of the wake radius from the local temporal mean, $R(x,t)-\overline{R}(x)$, comparing the envelope of the PIV data (grey region) with numerical solutions to the model (lines). The data envelope spans the minimum and maximum wake radii in the data over $t/T\in [0,1)$ at each streamwise location. Good agreement between the model solutions and measured data is observed for $u^*\leq 0.2$.}
\label{fig:envelopes}
\end{figure}

For a more direct comparison with the model results, the envelope of the wake-radius perturbation was computed from the data. This represents the minimum and maximum values observed in the quantity $R(x,t)-\overline{R}(x)$ over $t/T\in[0,1)$. The envelope representation removes the effects of wake recovery in terms of streamwise changes in both the time-averaged radius $\overline{R}(x)$ and the wave advection velocity $\overline{U}(x)$, neither of which are captured in the model. Plots of these envelopes along with the model solutions at two representative time instances are shown in Figure \ref{fig:envelopes} for all four unsteady cases with $\lambda_0=4.50$. Given the simplifying assumptions of the modeling framework, the agreement between the model solutions and experimental results is remarkable. For the lower-amplitude cases ($u^*\leq 0.2$), the model shows a very similar rate of growth in the wake-radius amplitude as in the envelope of the data. The saturated amplitude of the wake-radius waves downstream of $x/D\approx 5$ also corresponds well between the model and data. The model overestimates the final amplitude of the wake radius in the case with the highest forcing amplitude ($u^*=0.4$), suggesting that the physics neglected in the modeling approach become more significant as the unsteady forcing amplitude increases.

\begin{figure}
\begin{subfigure}[t]{0.48\textwidth}
\centering
  \includegraphics[width=\textwidth]{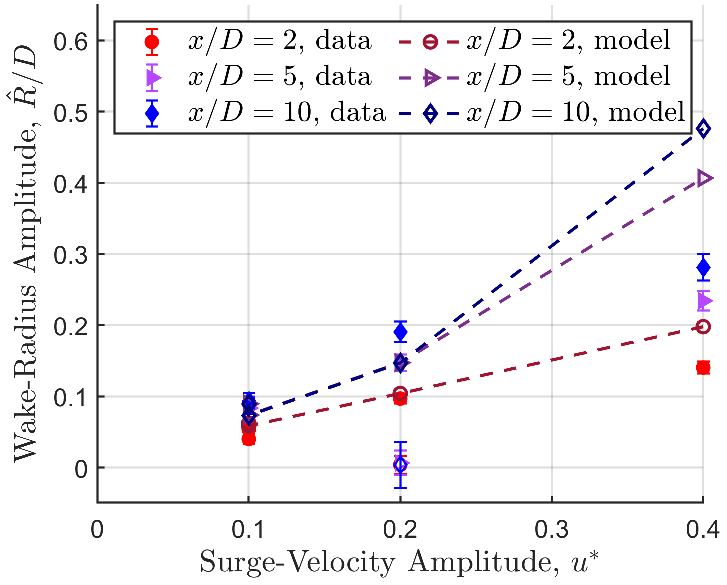}
  \caption{}
\label{fig:rAmps_TSRlo}
\end{subfigure}
\hfill
\begin{subfigure}[t]{0.48\textwidth}
\centering
  \includegraphics[width=\textwidth]{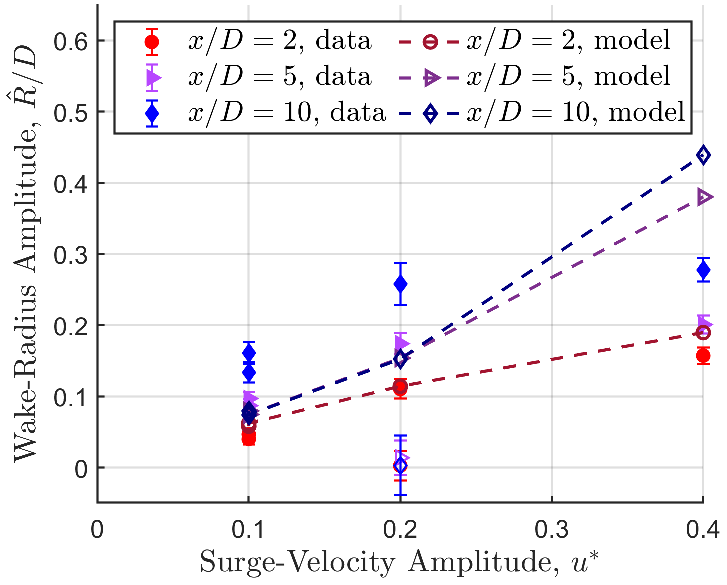}
  \caption{}
\label{fig:rAmps_TSRhi}
\end{subfigure}
\caption{Comparison between theoretical and experimental results for the wake-radius amplitude $\hat{R}/D$, plotted as a function of surge-velocity amplitude $u^*$. Experimental data are shown as colored markers, model solutions are shown as darker-colored open markers, and results obtained from quasi-steady measurements are given as colored open markers. Linear interpolations between the model solutions are shown as dashed lines. The colors and shapes of the markers correspond to their streamwise locations in the wake. Relatively good agreement between the model and data is observed for $x/D\leq5$ and $u^*\leq0.2$.}
\label{fig:rAmps}
\end{figure}

To further investigate the correspondence between the model results and experimental findings, we compute the amplitude of the wake-radius perturbations for both the model and the data for all cases. These are shown for both loading conditions in Figure \ref{fig:rAmps} for three streamwise locations: $x/D = 2$, 5, and 10, which represent the near-, intermediate-, and far-wake regions, respectively. The effective wake-radius amplitudes from the quasi-steady measurements at $U_\infty = 0.8$ and 1.2 $\rm{ms^{-1}}$ are shown as open colored markers at $u^*=0.2$. It is apparent that the quasi-steady approximation of the wake-radius amplitude does not capture the unsteady dynamics of the wake, as there are no traveling waves present in the quasi-steady approximation. The model results, shown as darker-colored open markers (with dashed lines representing linear interpolations between solutions), follow the trends in the data (closed markers) more closely. At the lower surge-velocity amplitudes ($u^*\leq0.2$) and in the near- and intermediate-wake regions, the model results align well with the data. The model solutions diverge from the data at the highest surge-velocity amplitude and in the far wake, as expected given the breakdown of its assumptions in these regimes. Still, even in these situations the model correctly predicts that the wake-radius amplitude will increase with increasing $u^*$ and with increasing $x/D$, suggesting that the underlying physics parameterized by the model are still relevant.

Results from a similar analysis of the amplitude of the streamwise velocity are shown in Figure \ref{fig:uAmps} in Appendix \ref{app:QS}. The model solutions strongly underpredicted these amplitudes but still captured the trends in the data, especially compared to the quasi-steady approximations. While the results suggest that the model in its current form is not a quantitatively accurate predictive tool for all wake properties, it is still able to represent key dynamical phenomena of these unsteady wake scenarios.


\subsection{Wake-recovery enhancements via unsteady flow mechanisms}\label{sec:results_recovery}

The preceding sections have demonstrated that the theoretical considerations in Section \ref{sec:theory} are able to qualitatively describe the growth and propagation of traveling-wave perturbations in unsteady turbine wakes, but are not able to directly address the streamwise evolution of time-averaged quantities. Using the experimental results, however, it is possible to observe the effects of unsteady wake forcing on these time-averaged quantities and consider their underlying physics.

\begin{figure}
\centering
\begin{subfigure}[t]{0.48\textwidth}
\centering
  \includegraphics[width=\textwidth]{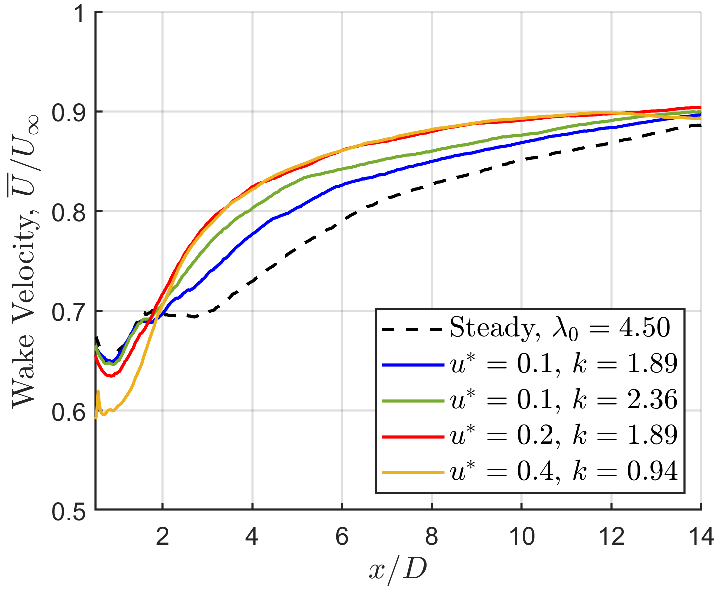}
  \caption{}
\label{fig:recovery_TSRlo}
\end{subfigure}
\hfill
\begin{subfigure}[t]{0.48\textwidth}
\centering
  \includegraphics[width=\textwidth]{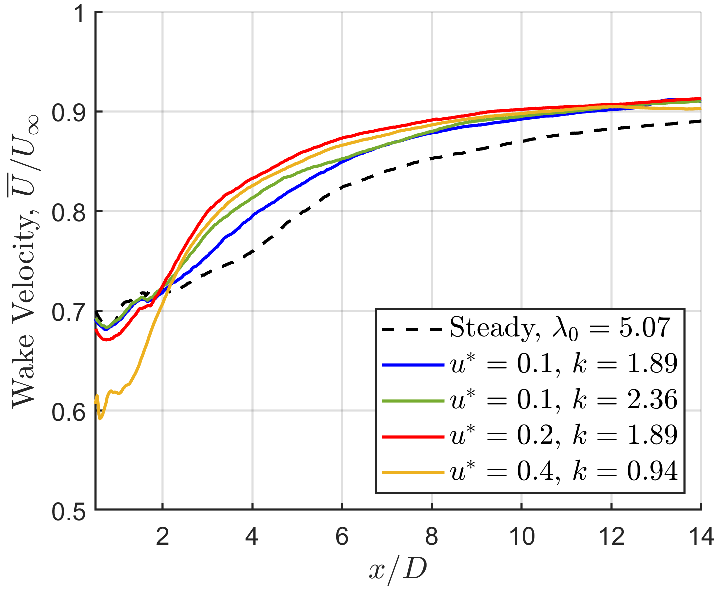}
  \caption{}
\label{fig:recovery_TSRhi}
\end{subfigure}
\caption{Radial- and time-averaged streamwise velocity $\overline{U}$ in the wake of the turbine for $\lambda_0=4.50$ (a) and $\lambda_0=5.07$ (b). The steady reference cases are shown as black dashed lines.}
\label{fig:wake_recovery}
\end{figure}

The radial- and time-averaged streamwise velocity $\overline{U}(x)$ is shown for all cases in Figure \ref{fig:wake_recovery}. The steady-flow cases, shown as black dashed lines, lie below the unsteady cases from $x/D\approx 2$ and deep into the far wake ($x/D>14$). The gap between the steady and unsteady wake profiles is most apparent in the intermediate wake, around $2\lesssim x/D \lesssim 6$, which is precisely where traveling-wave growth and tip-vortex aggregation occur. The extent to which the wake recovery is enhanced also increases with increasing surge-velocity amplitude as well as reduced frequency.




\begin{figure}
\centering
\begin{subfigure}[t]{0.32\textwidth}
\centering
  \includegraphics[width=\textwidth]{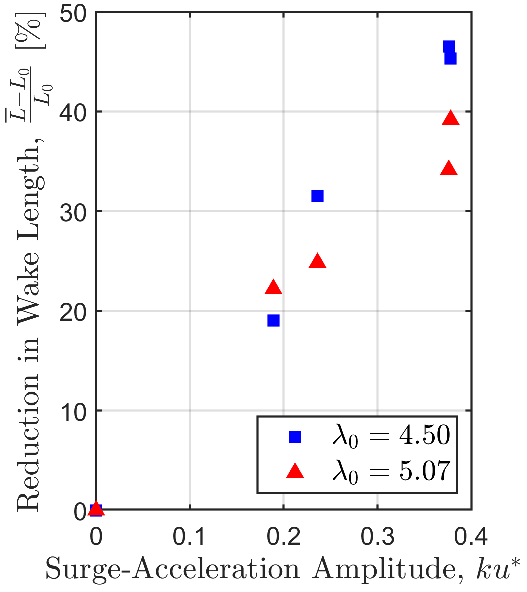}
  \caption{}
\label{fig:recstat_L}
\end{subfigure}
\hfill
\begin{subfigure}[t]{0.32\textwidth}
\centering
  \includegraphics[width=\textwidth]{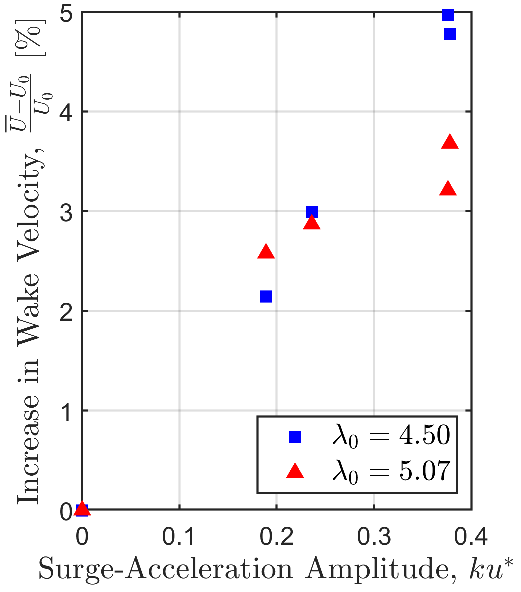}
  \caption{}
\label{fig:recstat_U}
\end{subfigure}
\hfill
\begin{subfigure}[t]{0.32\textwidth}
\centering
  \includegraphics[width=\textwidth]{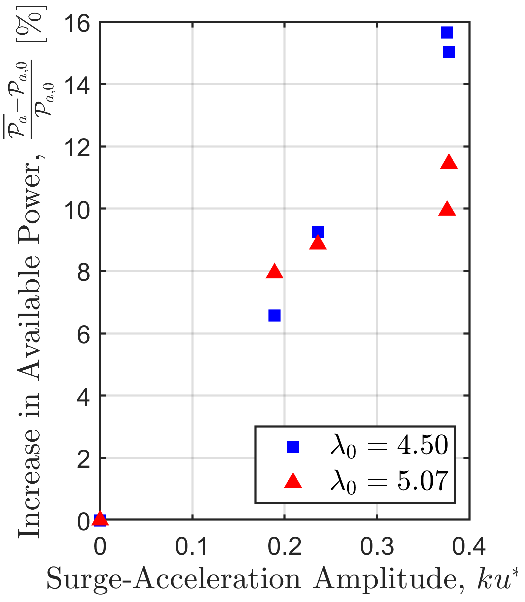}
  \caption{}
\label{fig:recstat_P}
\end{subfigure}
\caption{Measures of wake-recovery enhancement, reported as percentages. (a) shows the reduction in the streamwise distance $\overline{L}$ required for the wake to recover to $U=0.85U_\infty$, relative to the distance in the steady case $L_0$. (b) shows the enhancement in the streamwise wake velocity $\overline{U}$ at $x/D = 10$, again relative to the corresponding steady-flow quantity $U_0(x/D=10)$. (c) shows the enhancement in the power available in the wake at $x/D = 10$ relative to the steady case, $\left(\overline{\mathcal{P}_a}-\mathcal{P}_{a,0}\right)/\mathcal{P}_{a,0}$.}
\label{fig:recovery_stats}
\end{figure}

The velocity profiles in Figure \ref{fig:wake_recovery} can be used to further quantify the degree of wake enhancement observed in these experiments. First, a reduction in the streamwise extent of the wake can be calculated by defining the time-averaged streamwise distance $x=\overline{L}$ required for the wake velocity to recover to $U(x)=0.85U_\infty$. The percent reduction in this wake-length variable is shown in Figure \ref{fig:recstat_L}. All unsteady cases showed reductions in $\overline{L}$ by at least 19\%, with improvements in excess of 45\% achieved at the highest amplitudes. Second, a similar quantity can be computed for the increase in wake velocity at a fixed streamwise distance of $x/D=10$, a reasonable inter-turbine spacing for a moderate-sized wind farm \citep{stevens_flow_2017}. As shown in Figure \ref{fig:recstat_U}, the wake velocities were increased at this particular location in all unsteady cases by 2 to 5\% relative to the steady case. Taking the cube of these improvements gives a measure of the increase in the available power in the flow at the same streamwise location, shown in Figure \ref{fig:recstat_P}. In these experiments, enhancements in the available power in the flow at $x/D=10$ ranged from 6.7 to 15.7\%, which could represent a large potential increase in power generation for a downstream turbine placed at this location in an array. These findings are in line with the recent experimental results of \cite{bossuyt_floating_2023} and \cite{messmer_enhanced_2024} for FOWTs, as well as those of \cite{frederik_periodic_2020} and \cite{van_der_hoek_experimental_2022} for fixed turbines under dynamic induction control.

The various measures of wake-recovery enhancement in Figure \ref{fig:recovery_stats} are plotted against the surge-acceleration amplitude $ku^*$. This choice leads to a better collapse in the data than plotting against $u^*$ or $k$ alone. The data from the lower tip-speed ratio cases seem to scale in direct proportionality with the surge acceleration, while the data from the higher tip-speed ratio cases appear to saturate toward higher acceleration amplitudes. This scaling may be a result of the time-varying thrust force from the surging turbine that generates the unsteady wake dynamics. The thrust force exerted by the turbine on the flow can be approximated as the sum of quasi-steady and added-mass forces, 

\begin{equation}
    F_T \sim \frac{\pi}{8} \rho D^2 C_T (U_\infty - \mathcal{U}(t))^2 + \frac{1}{2} \rho V_d C_a \frac{\partial \mathcal{U}}{\partial t}\,,
    \label{eqn:thrust}
\end{equation}

\noindent where $C_T$ is the thrust coefficient, $V_d$ is the effective volume of fluid displaced by the rotor disc, and $C_a$ is an added-mass coefficient. Assuming that $C_T$ and $C_a$ do not vary appreciably with time, the nondimensional force amplitude applied by the turbine on the flow in the wake scales as 

\begin{equation}
    \frac{\hat{F}}{F_0} \sim u^* + ku^*
    \label{eqn:thrust_scaling}
\end{equation}

\noindent to first order. The fact that the data in Figure \ref{fig:recovery_stats} appear to scale with $ku^*$ may suggest that the flow accelerations associated with the effective added mass of the surging turbine represent an additional unsteady forcing on the wake that aids in the recovery of kinetic energy. In effect, the surge motions of the turbine may be pumping the flow in the wake, thereby stimulating faster wake recovery.

\begin{figure}
\begin{subfigure}[t]{0.48\textwidth}
\centering
  \includegraphics[width=\textwidth]{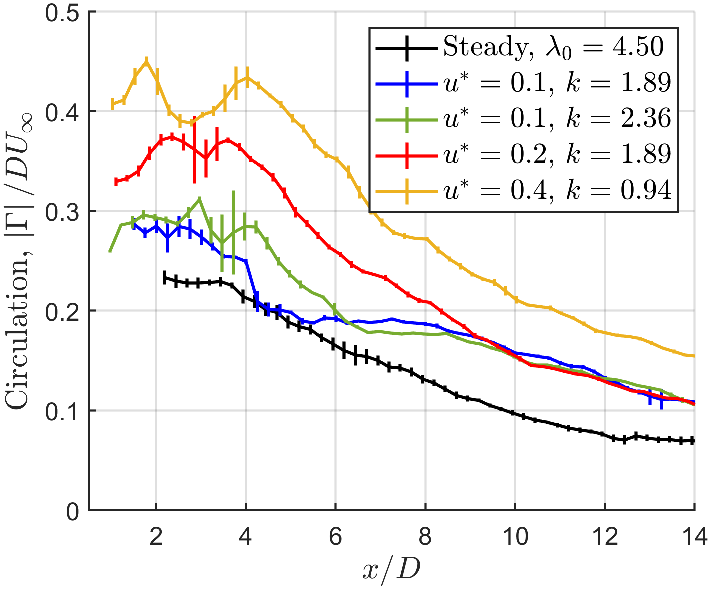}
  \caption{}
\label{fig:circ_TSRlo}
\end{subfigure}
\hfill
\begin{subfigure}[t]{0.48\textwidth}
\centering
  \includegraphics[width=\textwidth]{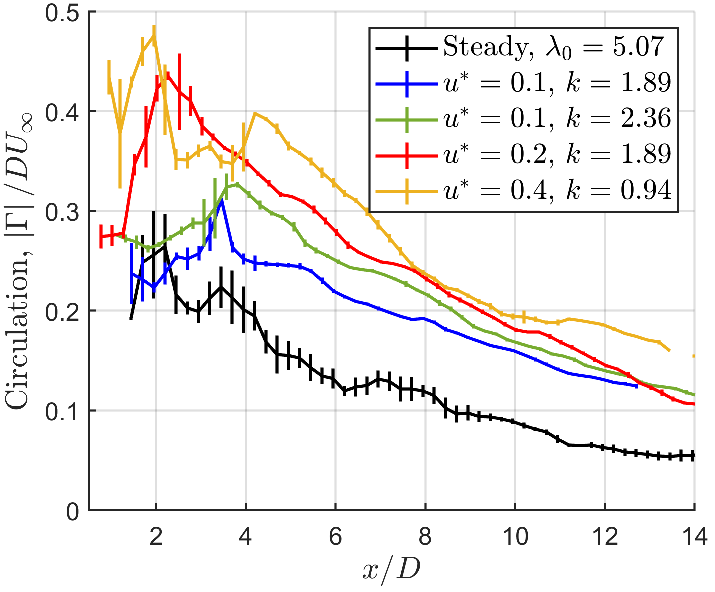}
  \caption{}
\label{fig:circ_TSRhi}
\end{subfigure}
\caption{Circulations of vortex-aggregate structures in the wake, averaged over a surge period in streamwise bins of width $D/4$. The steady reference case is shown in black.}
\label{fig:circulations}
\end{figure}

Complementary wake-recovery mechanisms may be postulated to explain the observed enhancements. For instance, the vortex aggregates visible in the intermediate and far wake in Figures \ref{fig:example_vort} and \ref{fig:cases_vort} likely enhance mixing and the entrainment of free-stream momentum into the wake. To examine this hypothesis, the centroids of these structures were identified in the vorticity fields, the circulations were computed on a circular path of radius $D/4$ around the centroids, and the circulations were bin-averaged by streamwise distance over all time instances for each case. These circulations are shown in Figure \ref{fig:circulations} for both loading conditions, and the data show that increasing the surge-velocity amplitude and surge frequency increase the circulations of the vortex aggregates throughout the wake. This aligns well with the idea that the surge motions of the turbine are applying a pulsatile forcing on the wake, since the roll-up of these structures can be described by the 1D modeling framework, and those dynamics are driven by the forcing amplitude of the wake velocity at the upstream boundary.

The geometry of the wake in the unsteady cases may also play a role in wake-recovery enhancement. The wavy wake shape seen in Figures \ref{fig:example_wakeRadius} and \ref{fig:r_vs_x_data} implies that the unsteady wakes have a larger surface area than their steady-flow counterparts, which may give rise to additional turbulent momentum transport due to the larger interface. Similarly, the radial deformations in the wake themselves may encourage momentum transport in the radial direction. Though the 1D modeling framework ignores the radial velocity component in the wake, in reality an increase in the wake radius corresponds to a net momentum flux in the positive radial direction that scales with $U \frac{\partial R}{\partial t}$. This local bulk flow in the radial direction may encourage sweeps of high-momentum fluid into the wake or ejections of low-momentum fluid out of the wake, depending on the direction of the local radial flow.

In practice, the thrust-amplitude, vortex-aggregate, and wake-geometry contributions to wake recovery are most likely all coupled, as the thrust-force variations drive both traveling waves in the wake and tip-vortex roll-up. A more detailed analysis of momentum transport across the wake is not possible here due to insufficient ensemble sizes for resolving turbulent fluxes. Future work could investigate these mechanisms more directly and parameterize their effects on wake recovery.

\subsection{Discussion and implications}
\label{sec:discussion}

The experimental results described in the previous sections highlight several strengths of the modeling approach, as the theoretical hypotheses listed at the end of Section \ref{sec:theory} are confirmed in the data. Traveling waves are observed in the streamwise velocity and undergo damping as they propagate downstream, though the rate of this decay is not quantitatively captured by the model. The model also predicts the traveling-wave dynamics evident in the wake-radius measurements, including wave steepening and amplitude saturation. The tip-vortex aggregation explored in Section \ref{sec:theory_vortex} is observed in the measured vorticity fields as well. Finally, the strengths of these unsteady dynamical features all increase with increasing forcing amplitude, represented by higher-amplitude oscillations at the inlet to the wake that scale with increasing surge-velocity amplitudes of the turbine.

The experimental results also highlight the assumptions and limitations of the theoretical analysis. Since the modeling approach in its current form neglects pressure gradients and wake-recovery effects, it is unable to capture changes in time-averaged quantities as a function of downstream distance. Mean-flow effects, such as pressure recovery and turbulent entrainment of momentum from the free stream into the wake, will alter the propagation velocity of traveling waves in the wake, the evolution of streamwise-velocity gradients, and ultimately the growth and saturation of the wake radius. In a strict sense, then, the model can only be used to predict perturbations about time-averaged quantities, and an extension would be required to capture the effects of wake recovery on the streamwise development of the time-averaged quantities themselves. These constraints are especially evident in the far-wake region where turbulent momentum entrainment contributes more significantly to wake recovery, as well as at high forcing amplitudes (e.g.\ $u^*=0.4$) where additional nonlinearities not included in the model become more significant. Despite these limiting assumptions, the dynamics described by the model are observed to correlate with enhancements in wake recovery, suggesting that these unsteady mechanisms strongly influence momentum transport into the wake.

Additional limitations of the present approach suggest possibilities for future theoretical work. First, the relationship between the turbine dynamics (including rotor inertia, induction, thrust, and power) and the inlet boundary condition of the model is not explicitly defined. This was done intentionally to preserve the general applicability of the framework, but an analytical connection could be derived by extending the framework of \cite{wei_power-generation_2023} to include near-wake quantities, as done by \cite{heck_modelling_2023} for yaw-misaligned turbines. Such an approach could allow other effects to be included as well, including yaw and tilt misalignment or aeroelastic turbine-blade deformations \citep[cf.][]{rodriguez_strongly-coupled_2020}. Also, a more physically motivated model for the streamwise-velocity autocovariance $\widetilde{U'U'}$ than the empirical linear fit used in this work could be implemented. Furthermore, the model could account for the pressure oscillations that occur downstream of turbines with time-varying thrust-force profiles. Finally, the vortex-dynamics analysis in Section \ref{sec:theory_vortex} only considers one-way coupling from the wake properties to the induced motions of the vortex elements. The effects of the vortices on the flow field in the wake have not been considered, primarily because the treatment of the wake dynamics has been limited to 1D while point vortices generate velocity components in two dimensions.

The experimental approach taken in this study also involves necessary abstractions that limit the direct applicability of these findings to utility-scale wind farms. The Reynolds number regime in the experiments is two orders of magnitude lower than that of full-scale wind turbines, which affects both the turbine aerodynamics \citep{miller_horizontal_2019} and the wake properties \citep{pique_laboratory_2022}. However, since the theoretical framework neglects viscous stresses in the wake, the mechanisms identified in the study should still be representative of those that would dominate the flow physics of full-scale rotors. The experiments also only considered a narrow range of tip-speed ratios and did not measure thrust forces on the turbine, which limit the conclusions that can be drawn from these data regarding turbine aerodynamics and power generation.

The most significant limitation of this study regarding its applicability to real-world flow scenarios is the lack of inflow turbulence in the towing tank. Since the flow in the tank was nominally quiescent before each run of the traverse, background turbulence in the facility was minimal and wake recovery in the baseline case was slow relative to turbine wakes in flows with higher turbulence intensities \citep[cf.][]{duckworth_investigation_2008}. This could suggest that the relatively large enhancements in wake recovery due to the surge motions of the turbine may not be as significant in atmospheric boundary-layer conditions, where the additional mixing provided by the unsteady wake dynamics may be overshadowed by the mixing from the highly turbulent background flow. Experimental results in agreement with this hypothesis have been reported by \cite{kadum_wind_2019}.

With these considerations in mind, the theoretical framework and experimental findings presented here still have important implications for the dynamics and control of utility-scale wind turbines and wind farms. For onshore and fixed-bottom offshore wind turbines, the unsteady wake dynamics explored in this study describe the effects of dynamic induction control on the wake. Changing the collective pitch angle of the turbine blades or the tip-speed ratio of the turbine in a periodic manner will produce a periodic variation in the turbine thrust force, thus exciting oscillations in streamwise velocity that lead to traveling-wave propagation, tip-vortex aggregation, and accelerated wake recovery. In fact, similar dynamics have been reported in previous studies on dynamic induction control. For instance, the vortical structures observed in the large-eddy simulations of \cite{munters_towards_2018} may be analogous to the vortex aggregates found in the present study. Similarly, the 3D flow measurements of \cite{van_der_hoek_experimental_2022} in the wake of a turbine with periodic variations in collective blade pitch also show propagating oscillations in the streamwise velocity in the near wake, as well as undulatory deformations in the wake radius and changes in the tip-vortex dynamics. It is noteworthy that both of these studies involve background flows with relatively high turbulence intensities and still show qualitatively similar dynamics to the present findings. This supports the broader applicability of the present work to utility-scale wind farms in more realistic atmospheric conditions.

The results of this study also highlight the potential benefits of unsteady turbine motion for wake-recovery enhancement. The typical methods for dynamic induction control on stationary turbines are limited by the thrust variations that can be achieved by changing the blade pitch angle and rotation rate. Surging the turbine back and forth can lead to much larger thrust-force variations, since the thrust force scales with the square of the rotor-frame velocity $U_\infty - \mathcal{U}(t)$ \citep{johlas_floating_2021,wei_power-generation_2023}. The unsteady platform motions of floating offshore wind turbines could thus be leveraged to achieve faster wake recovery. This has recently been corroborated in the findings of \cite{messmer_enhanced_2024}. These same unsteady motions can also be exploited to enhance the power generation of the surging turbine itself \citep{wei_phase-averaged_2022,wei_power-generation_2023}. Therefore, the unsteady dynamics associated with the natural rocking motions of floating offshore wind turbines can be harnessed to achieve higher power densities in floating offshore wind farms relative to their fixed-bottom counterparts. The enhancements in available power shown in Figure \ref{fig:recstat_P}, taken together with the power-generation improvements of up to 6.4\% above the stationary-turbine case reported by \cite{wei_phase-averaged_2022}, suggest that these increases in power density could be on the order of ten percent over conventional wind farms. It is also important to note that the power-density benefits in the current and aforementioned experimental studies were achieved without any active turbine control or array-scale optimization. Thus, even greater enhancements may be possible with the addition of physics-informed control strategies, though attention must be paid to the dynamic coupling between the controller, turbine thrust force, and floating-platform hydrodynamics \citep{van_den_berg_dynamic_2023}.

\section{Conclusions}\label{sec:conclusions}

This study demonstrates the effects of unsteady streamwise motions of a turbine on wake dynamics and recovery. A theoretical model of the wake radius and streamwise velocity in a wake with an oscillatory upstream boundary condition describes mechanisms for the growth and propagation of traveling waves in the wake, which encourage the roll-up of tip vortices into periodically shed vortical aggregates that advect into the far wake. These dynamics scale with the amplitude of the turbine motions. Experimental measurements in a towing-tank facility confirm the qualitative trends parameterized by the theoretical approach, and additionally show enhancements in wake recovery and the corresponding power available in the flow downstream of the turbine. These results support the findings of prior investigations involving dynamic induction control strategies for improving wind-farm power density. Taken together with power-generation enhancements observed in periodically surging turbines, these flow physics could yield improvements on the order of ten percent in floating offshore wind farms, relative to their fixed-bottom counterparts. The theoretical approach and findings of this study can also be applied to dynamic induction control strategies for traditional onshore and fixed-bottom offshore turbines in unsteady streamwise flow conditions such as axial gusts, as well as the dynamics and control of hydrokinetic turbine arrays in streamwise oscillatory tidal flows. Future work will investigate the coupled effects of floating offshore platform motions and turbine control schemes on wake recovery, as well as the unsteady loading effect of the traveling-wave disturbances on downstream turbines in an array.

\backsection[Acknowledgements]{The authors would like to thank Pengming Guo for help with setting up the experiments. The derivation of the theoretical framework was aided by insightful discussions with Omkar B.\ Shende, Young R.\ (Paul) Yi, and Lena Sabidussi. A colormap from the \textit{cmocean} library \citep{thyng_true_2016} was used for Figures \ref{fig:example_u} and \ref{fig:cases_u}.}

\backsection[Funding]{This work was supported by the National Science Foundation (grant number CBET-2038071) as well as the Natural Sciences and Engineering Research Council of Canada (grant number RGPIN-2023-03525). N.J.W. was supported by a National Science Foundation Graduate Research Fellowship and a Distinguished Postdoctoral Fellowship from the Andlinger Center for Energy and the Environment at Princeton University.}

\backsection[Declaration of interests]{The authors report no conflict of interest.}

\backsection[Data availability statement]{The data that support the findings of this study are available upon reasonable request.}

\backsection[Author ORCID]{N.\ J.\ Wei, \href{https://orcid.org/0000-0001-5846-6485}{https://orcid.org/0000-0001-5846-6485}; A.\ El Makdah, \href{https://orcid.org/0000-0003-4520-6259}{https://orcid.org/0000-0003-4520-6259}; F.\ Kaiser, \href{https://orcid.org/0000-0001-9888-8770}{https://orcid.org/0000-0001-9888-8770}; D.\ E.\ Rival, \href{https://orcid.org/0000-0001-7561-6211}{https://orcid.org/0000-0001-7561-6211}; J.\ O.\ Dabiri, \href{https://orcid.org/0000-0002-6722-9008}{https://orcid.org/0000-0002-6722-9008}}

\backsection[Author contributions]{N.J.W. derived the theory and analyzed the data. N.J.W., A.E.M., J.C.H., and F.K. performed the experiments. D.E.R. and J.O.D. procured funding and provided guidance on the project direction. All authors contributed to reaching conclusions, as well as writing and revising the paper.}

\FloatBarrier

\appendix
\section{Derivation of the governing equations}\label{app:derivations}

\begin{figure}
\centering
  \includegraphics[width=0.6\textwidth]{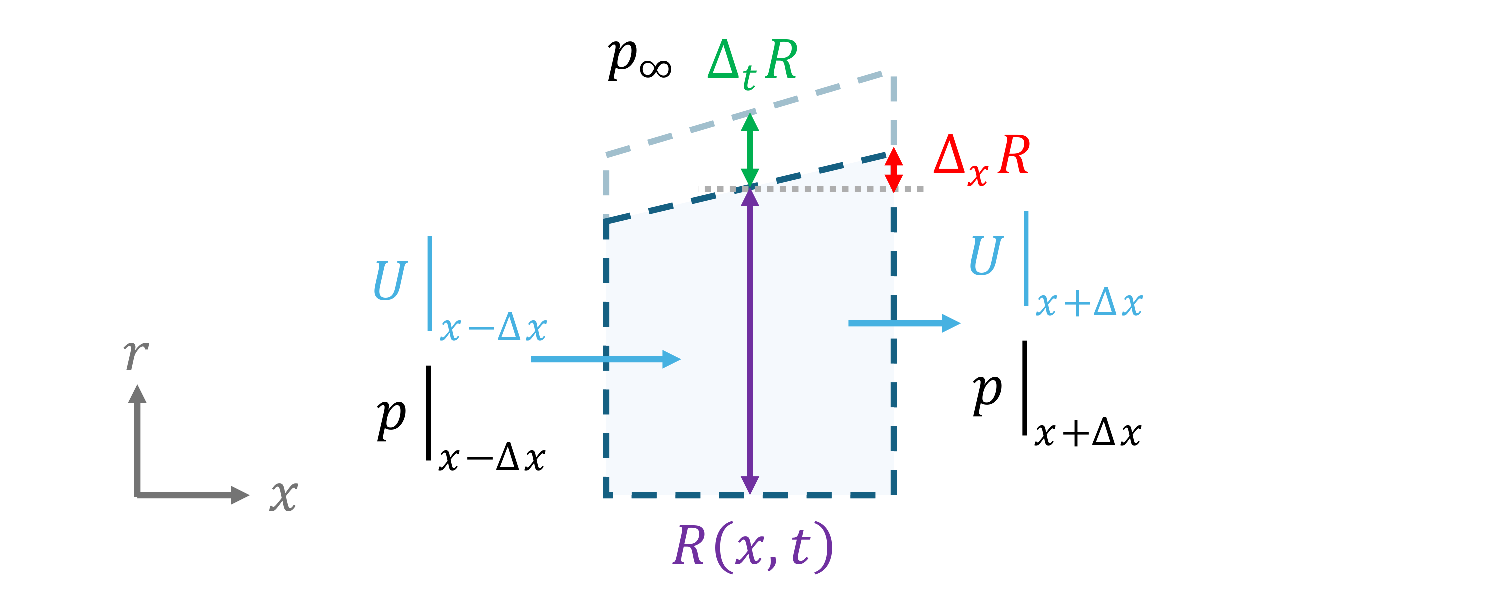}
  \caption{Sketch of the differential control volume used to derive conservation relations for an unsteady, radially deforming wake.}
\label{fig:diffCV}
\end{figure}

We consider incompressible and inviscid flow through a differential streamwise slice of the larger wake control volume shown in Figure \ref{fig:CV}, with streamwise width $2\Delta x$ and central radius $R(x,t)$. The wake radius is assumed to vary linearly with streamwise position and time, with spatial variation $\Delta_x R$ over a streamwise distance of $\Delta x$ and a temporal variation $\Delta_t R$ over a time step of $\Delta t$. The streamwise velocity through the differential control volume is $U|_{x-\Delta x}$ at the inlet and $U|_{x+\Delta x}$ at the outlet. The pressure outside the radius of the control volume is denoted $p_\infty$ and may vary with space and time. Radial velocities are neglected, as these will generally be smaller than the streamwise velocities through the control volume. The outer radius of the control volume has zero flow across its surface. A sketch of this differential control volume is given in Figure \ref{fig:diffCV}.

First, we derive the relation for conservation of mass, given in final form as Equation \ref{eqn:continuity}. The integral form of mass conservation for the differential control volume is

\begin{equation}
    \frac{\partial}{\partial t} \int_{cv} \rho\,dV + \int_{cs} \rho U\,dA = 0\,,
    \label{eqn:app_cont_int}
\end{equation}

\noindent where $\int_{cv} dV$ denotes integration over the control volume, $\int_{cs} dA$ is a flux integral over the control surface, and $\rho$ is the fluid density. We use a second-order finite-difference approximation for the growth in the control volume as a function of time,

\begin{equation}
\frac{\partial V}{\partial t} \approx \frac{V|_{t+\Delta t}-V|_{t-\Delta t}}{2\Delta t}\,.
\end{equation}

\noindent The volume of the control volume can be calculated from the volume of a partial cone with height $h$ and radii $R_1$ and $R_2$,

\begin{equation}
    V = \frac{1}{3}\pi h \left(R_1^2 +R_1 R_2 + R_2^2\right)\,.
\end{equation}

\noindent We therefore obtain the unsteady volumetric growth in the control volume as 

\begin{equation}
    \frac{\partial V}{\partial t} \approx \frac{2\pi \Delta x \left(R^2 + 2R\Delta_t R\right) - 2\pi \Delta x \left(R^2 - 2R\Delta_t R\right)}{2\Delta t} = 4\pi R \Delta x \frac{\Delta_t R}{\Delta t}\,,
\end{equation}

\noindent where we have neglected terms of order $\Delta^2$ and above. Including the flux terms, Equation \ref{eqn:app_cont_int} becomes

\begin{multline}
    \frac{\partial}{\partial t} \int_{cv} \,dV + \int_{cs}  U\,dA \\
    = 4\pi R \Delta x \frac{\Delta_t R}{\Delta t} + \pi U|_{x+\Delta x}\left(R+\Delta_x R\right)^2 - \pi U|_{x-\Delta x}\left(R-\Delta_x R\right)^2 \\
    \approx 4\pi R \Delta x \frac{\Delta_t R}{\Delta t} + \pi U|_{x+\Delta x}\left(R^2+\Delta_x R\right) - \pi U|_{x-\Delta x}\left(R^2-\Delta_x R\right) = 0\,,
\end{multline}

\noindent again neglecting higher-order products of $\Delta$. Assuming $U$ varies linearly with $x$, we can write $U|_{x+\Delta x} + U|_{x-\Delta x} = 2 U$ and $U|_{x+\Delta x} - U|_{x-\Delta x} = 2 \Delta_x U$. Simplifying the relation for continuity, we obtain

\begin{equation}
    4 R \Delta x \frac{\Delta_t R}{\Delta t} + 2 R^2 \Delta_x U + 4RU \Delta_x R = 0\,.
\end{equation}

\noindent Dividing by $4r\Delta x$ and taking the limits as $\Delta x\rightarrow 0$ and $\Delta t \rightarrow 0$, we arrive at the differential form of the continuity equation,

\begin{equation}
    \frac{\partial R}{\partial t} + \frac{1}{2} R \frac{\partial U}{\partial x} + u \frac{\partial R}{\partial x} = 0\,,
    \label{eqn:app_cont}
\end{equation}

\noindent which corresponds to Equation \ref{eqn:continuity}.

A similar process can be undertaken to derive the differential form of the momentum equation (Equation \ref{eqn:momentum_Euler}). The integral form of conservation of momentum in the streamwise direction is

\begin{equation}
    \frac{\partial}{\partial t} \int_{cv} \rho U\,dV + \int_{cs} \rho U^2\,dA = -\int_{cs}p\,dA\,.
    \label{eqn:app_momt_int}
\end{equation}

\noindent The unsteady term can be integrated over the differential control volume using cylindrical coordinates:

\begin{multline}
    \frac{\partial}{\partial t} \int_{cv} U\,dV = 2\pi \int_{-\Delta x}^{\Delta x} \frac{\partial}{\partial t} \int_0^{R(t)} U r'\,dr' dx' 
    = 2\pi \int_{-\Delta x}^{\Delta x} \left[ UR\frac{\partial R}{\partial t} + \int_0^{R(t)} \frac{\partial}{\partial t} Ur'\,dr' \right] dx' \\
    \approx 2\pi \Delta x \left(2UR \frac{\partial R}{\partial t} + R^2 \frac{\partial U}{\partial t}\right) \,,
\end{multline}

\noindent employing the Leibniz integral rule and neglecting higher-order terms in the integration over $x'$. For the advection term, we have

\begin{multline}
    \int_{cs} U^2\,dA = {U|_{x+\Delta x}}^2 \pi \left(R+\Delta_x R\right)^2 - {U|_{x-\Delta x}}^2 \pi \left(R-\Delta_x R\right)^2 \\
     \approx \pi R^2 \left({U|_{x+\Delta x}}^2 - {U|_{x-\Delta x}}^2\right) + 2\pi R \Delta_x R \left({U|_{x+\Delta x}}^2 + {U|_{x-\Delta x}}^2\right) \\
     = \pi R^2 \left({U|_{x+\Delta x}} + {U|_{x-\Delta x}}\right) \left({U|_{x+\Delta x}} - {U|_{x-\Delta x}}\right) \\ + 2\pi R \Delta_x R \left(\left({U|_{x+\Delta x}} + {U|_{x-\Delta x}}\right)^2 - 2 U|_{x+\Delta x} U|_{x-\Delta x}\right) \\
     \approx 4\pi R U \left(R \Delta_x U + U \Delta_x R\right) \,,
\end{multline}

\noindent where we have neglected higher-order terms, completed the square, and noted that $U|_{x+\Delta x} U|_{x-\Delta x} = (U+\Delta U)(U-\Delta U) \approx U^2$. For a pressure distribution outside of the wake $p_{\infty}$ that does not vary in the radial direction, the pressure contribution to the momentum balance may be written as

\begin{multline}
    \int_{cs} p\,dA = \pi\left(R+\Delta_x R\right)^2 p|_{x+\Delta x} - \pi\left(R-\Delta_x R\right)^2 p|_{x-\Delta x} - 2\pi \int_{R-\Delta_x R}^{R+\Delta_x R} r' p_{\infty}\,dr' \\
    \approx \pi\left(R^2+2r\Delta_x R\right) p|_{x+\Delta x} - \pi\left(R^2-2r\Delta_x R\right)p|_{x-\Delta x} - 4\pi p_{\infty} R \Delta_x R \\
    = 2\pi R \left(R \Delta_x p + 2 (p-p_{\infty}) \Delta_x R\right)  \,.
\end{multline}

\noindent Finally, putting all of these terms together, the momentum equation over the differential control volume is given by 

\begin{equation}
    2\pi \Delta x \left(2UR \frac{\partial R}{\partial t} + R^2 \frac{\partial U}{\partial t}\right) + 4\pi R U \left(R\Delta_x U + U \Delta_x R\right) = 2\pi R \left(R\Delta_x p + 2(p-p_{\infty})\Delta_x R\right)\,
\end{equation}

\noindent which, after dividing by $2\pi R \Delta x$ and taking the limit as $\Delta x \rightarrow 0$, yields

\begin{equation}
    R \frac{\partial U}{\partial t} + 2U\left(\frac{\partial R}{\partial t} + R \frac{\partial U}{\partial x} + U \frac{\partial R}{\partial x}\right) = -\frac{1}{\rho} \left(R \frac{\partial p}{\partial x} + 2 \frac{\partial R}{\partial x} (p-p_\infty)\right)\,.
\end{equation}

\noindent Subtracting Equation \ref{eqn:app_cont} from this equation gives

\begin{equation}
    \frac{\partial U}{\partial t} + U \frac{\partial U}{\partial x} = -\frac{1}{\rho} \frac{\partial p}{\partial x} - \frac{2}{\rho R} \frac{\partial R}{\partial x}\left(p-p_\infty\right)\,.
    \label{eqn:app_momt_pinfty}
\end{equation}

\noindent If we assume that the pressure outside the wake is radially homogeneous, i.e.\ $p_\infty=p(x)$, then the extra pressure term becomes zero and we arrive at the 1D Euler equation,

\begin{equation}
    \frac{\partial U}{\partial t} + U \frac{\partial U}{\partial x} = -\frac{1}{\rho} \frac{\partial p}{\partial x}\,,
    \label{eqn:app_momt}
\end{equation}

\noindent which corresponds to Equation \ref{eqn:momentum_Euler}.

\section{Further discussion of the quasi-steady wake approximation}
\label{app:QS}

\begin{figure}
\begin{subfigure}[t]{0.48\textwidth}
\centering
  \includegraphics[width=\textwidth]{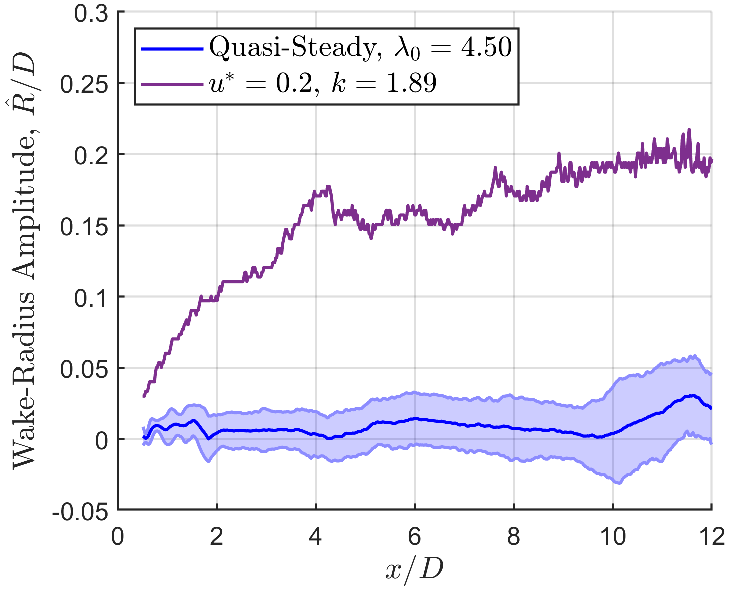}
  \caption{}
\label{fig:rAmp_QS_TSRlo}
\end{subfigure}
\hfill
\begin{subfigure}[t]{0.48\textwidth}
\centering
  \includegraphics[width=\textwidth]{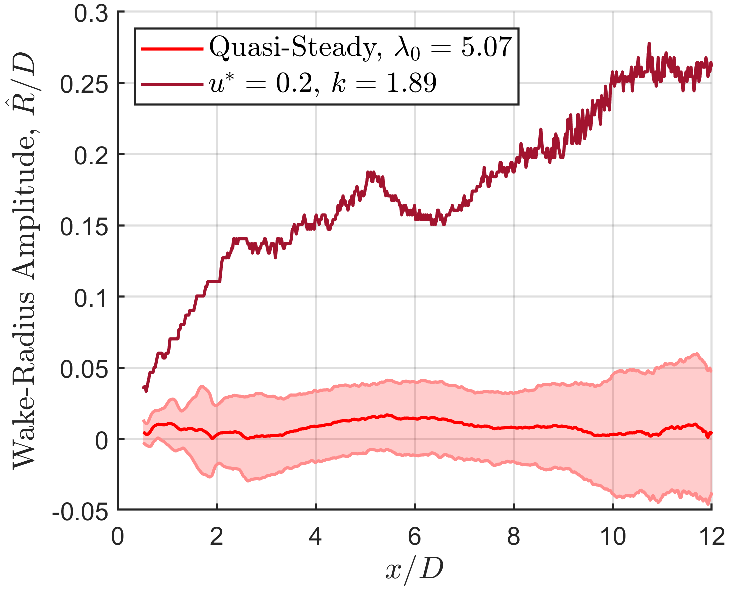}
  \caption{}
\label{fig:rAmp_QS_TSRhi}
\end{subfigure}
\caption{Amplitude of the wake radius, calculated from unsteady measurements with $u^*=0.2$ and $k=1.89$ (darker lines) and quasi-steady flow measurements at $U_\infty = 0.8$ and 1.2 $\rm{ms^{-1}}$ (lines with uncertainty bounds). Cases with $\lambda_0=4.50$ (a) and $\lambda_0=5.07$ (b) are shown.}
\label{fig:rAmp_QS}
\end{figure}

\begin{figure}
\begin{subfigure}[t]{0.48\textwidth}
\centering
  \includegraphics[width=\textwidth]{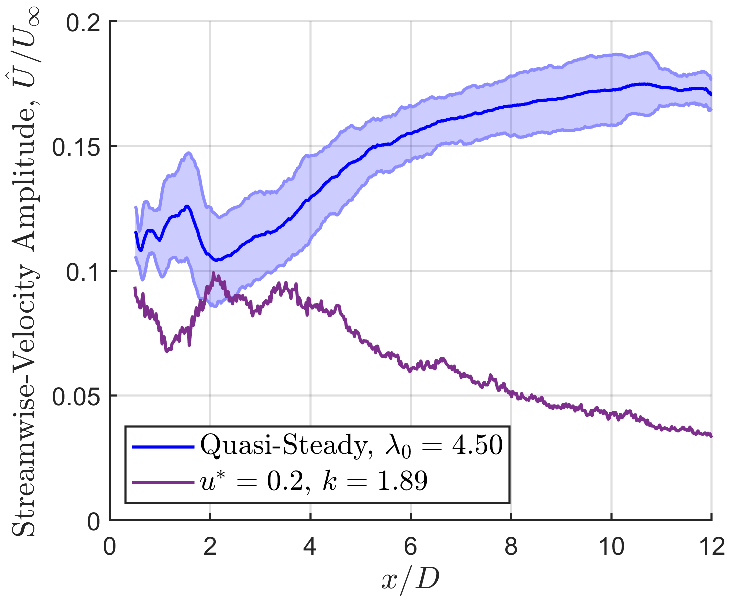}
  \caption{}
\label{fig:uAmp_QS_TSRlo}
\end{subfigure}
\hfill
\begin{subfigure}[t]{0.48\textwidth}
\centering
  \includegraphics[width=\textwidth]{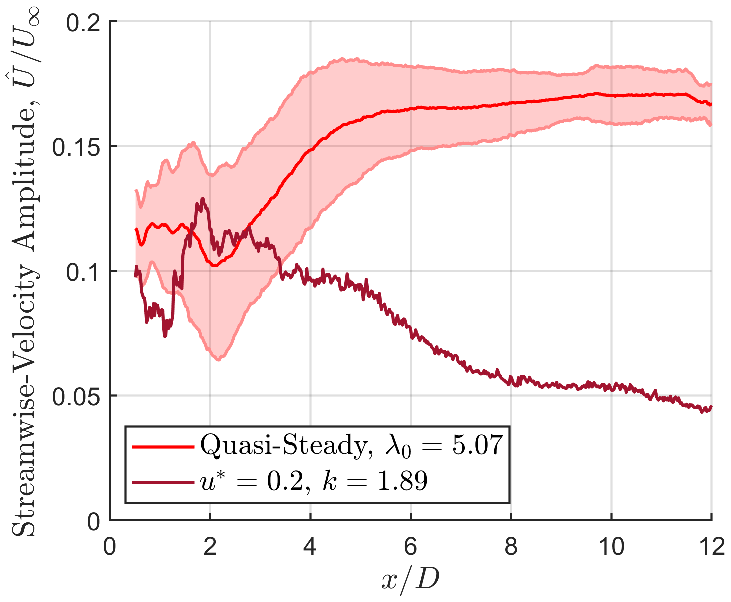}
  \caption{}
\label{fig:uAmp_QS_TSRhi}
\end{subfigure}
\caption{Amplitude of the streamwise velocity, calculated from unsteady measurements with $u^*=0.2$ and $k=1.89$ (darker lines) and quasi-steady flow measurements at $U_\infty = 0.8$ and 1.2 $\rm{ms^{-1}}$ (lines with uncertainty bounds). Cases with $\lambda_0=4.50$ (a) and $\lambda_0=5.07$ (b) are shown.}
\label{fig:uAmp_QS}
\end{figure}

The results presented in this study highlight the importance of unsteady fluid dynamics in the wakes of turbines with time-varying thrust profiles. The traveling-wave solutions and tip-vortex roll-up behaviors are not phenomena that can be obtained from quasi-steady flow assumptions. To demonstrate this even more clearly, the streamwise variations in the wake-radius amplitude $\hat{r}$ and streamwise-velocity amplitude $\hat{U}$ are shown in Figures \ref{fig:rAmp_QS} and \ref{fig:uAmp_QS} for both loading conditions. The dark-colored lines are taken from unsteady cases with $u^* = 0.2$ and $k = 1.89$, while the lines with shaded uncertainty intervals are calculated by subtracting the wake radius and streamwise velocity across steady-flow cases with $U_\infty = 0.8$ $\rm{ms^{-1}}$ and $U_\infty = 1.2$ $\rm{ms^{-1}}$ to represent an effective surge-velocity amplitude of $u^*=0.2$ (with $k=0$). The differences between the representative quasi-steady amplitudes and the measured unsteady amplitudes are clear. For the wake radius in the unsteady case, the amplitude increases with downstream distance. The differences in wake radius across the quasi-steady flow representations are small in comparison and do not display the same growth and saturation as seen in the measurements. Likewise, the streamwise-velocity amplitude shows opposite trends between the unsteady and quasi-steady measurements. While $\hat{U}$ decays with downstream distance in the unsteady case, in accordance with the modeling framework from Section \ref{sec:theory_eqns}, the quasi-steady amplitude grows. This is due to differences in the steady-flow velocity deficit between the two quasi-steady datasets. These comparisons further underscore the need for dynamic models of the flows in unsteady turbine wakes.

\begin{figure}
\begin{subfigure}[t]{0.48\textwidth}
\centering
  \includegraphics[width=\textwidth]{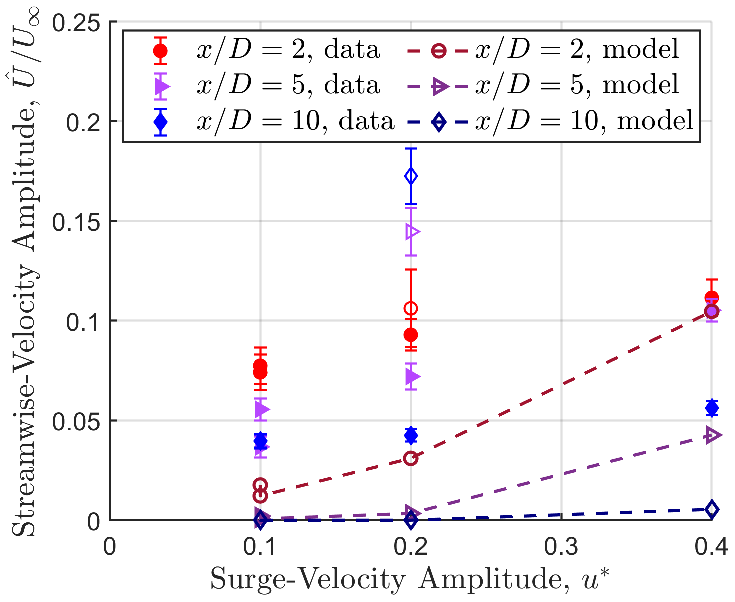}
  \caption{}
\label{fig:uAmps_TSRlo}
\end{subfigure}
\hfill
\begin{subfigure}[t]{0.48\textwidth}
\centering
  \includegraphics[width=\textwidth]{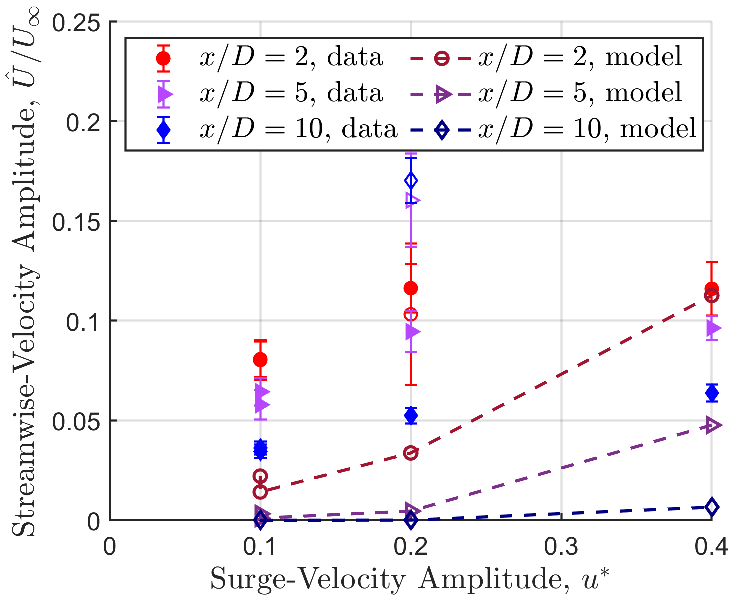}
  \caption{}
\label{fig:uAmps_TSRhi}
\end{subfigure}
\caption{Comparison between theoretical and experimental results for the amplitude of the streamwise velocity $\hat{U}/U_\infty$, plotted as a function of surge-velocity amplitude $u^*$. Experimental data are shown as colored markers, model solutions are shown as darker-colored open markers, and results obtained from quasi-steady measurements are given as colored open markers. Linear interpolations between the model solutions are shown as dashed lines. The colors and shapes of the markers correspond to their streamwise locations in the wake. The model shows some qualitative agreement with the trends observed in the data, but is not quantitatively accurate.}
\label{fig:uAmps}
\end{figure}

For the sake of completeness, we also plot the amplitudes of the radially averaged streamwise velocity as a function of surge-velocity amplitude, shown for both loading conditions in Figure \ref{fig:uAmps}. Unlike the wake-radius amplitudes shown previously in Figure \ref{fig:rAmps}, the streamwise-velocity amplitudes computed by the model strongly underpredict the measured data for most cases. This is likely due to the effects of wake recovery, which more directly impact the streamwise velocity than the wake radius. Despite this lack of quantitative agreement, the model still captures the trends in the data: the streamwise-velocity amplitude is shown to increase with increasing surge-velocity amplitude, and decrease with increasing streamwise distance. In accordance with the comparisons shown above in Figure \ref{fig:uAmp_QS}, the quasi-steady amplitudes show the opposite trend with streamwise distance. Though for this quantity the modeling approach does not yield accurate quantitative predictions, these results suggest that the unsteady flow physics it parameterizes are to some extent reflective of the full dynamics of unsteady turbine wakes.

\bibliographystyle{jfm}
\bibliography{Wakes}

\end{document}